\documentclass[fleqn,usenatbib]{mnras}

\usepackage{newtxtext,newtxmath}

\usepackage[T1]{fontenc}
\usepackage{ae,aecompl}

\usepackage[dvipdfmx]{graphicx}	
\usepackage{amsmath}	
\usepackage{threeparttable}
\usepackage{nccmath}
\usepackage{multirow}
\usepackage{url}
\usepackage{lineno} 

\newcommand*\patchAmsMathEnvironmentForLineno[1]{
  \expandafter\let\csname old#1\expandafter\endcsname\csname #1\endcsname
  \expandafter\let\csname oldend#1\expandafter\endcsname\csname end#1\endcsname
  \renewenvironment{#1}
     {\linenomath\csname old#1\endcsname}
     {\csname oldend#1\endcsname\endlinenomath}}
\newcommand*\patchBothAmsMathEnvironmentsForLineno[1]{
  \patchAmsMathEnvironmentForLineno{#1}
  \patchAmsMathEnvironmentForLineno{#1*}}
\AtBeginDocument{
\patchBothAmsMathEnvironmentsForLineno{equation}
\patchBothAmsMathEnvironmentsForLineno{align}
\patchBothAmsMathEnvironmentsForLineno{flalign}
\patchBothAmsMathEnvironmentsForLineno{alignat}
\patchBothAmsMathEnvironmentsForLineno{gather}
\patchBothAmsMathEnvironmentsForLineno{multline}
}

\title[Protocluster Cores at $1<z<1.5$]{A systematic search for galaxy protocluster cores at the transition epoch of their star formation activity}

\author[M. Ando et al.]{
Makoto Ando,$^{1}$\thanks{E-mail: mando@astron.s.u-tokyo.ac.jp}
Kazuhiro Shimasaku$^{1,2}$, Rieko Momose$^{3,1,4,5}$, Kei Ito$^{6,7}$, Marcin Sawicki$^{8}$,\newauthor and Rhythm Shimakawa$^{7}$ 
\\
$^{1}$Department of Astronomy, Graduate School of Science, The University of Tokyo, 7-3-1 Hongo, Bunkyo-ku, Tokyo 113-0033, Japan\\
$^{2}$Research Center for the Early Universe, The University of Tokyo, 7-3-1 Hongo, Bunkyo-ku, Tokyo 113-0033, Japan\\
$^{3}$Carnegie Observatories, 813 Santa Barbara Street, Pasadena, CA 91101, USA\\
$^{4}$Kavli Institute for the Physics and Mathematics of the Universe (WPI),
UTIAS, The University of Tokyo, Kashiwa, Chiba 277-8583, Japan\\
$^{5}$The Institute for AI and Beyond, The University of Tokyo, Tokyo
113-8655, Japan\\
$^{6}$Department of Astronomical Science, The Graduate University for Advanced Studies, SOKENDAI, 2-21-1 Osawa, Mitaka, Tokyo, 181-8588, Japan\\
$^{7}$National Astronomical Observatory of Japan (NAOJ), National Institutes of Natural Sciences, 2-21-1 Osawa, Mitaka, Tokyo 181-8588, Japan\\
$^{8}$Department of Astronomy and Physics and Institute for Computational Astrophysics, Saint Mary's University, 923 Robie Street, Halifax,\\  Nova Scotia B3H 3C 3, Canada
}

\date{Accepted XXX. Received YYY; in original form ZZZ}

\pubyear{2022}

\begin{document}
\label{firstpage}
\pagerange{\pageref{firstpage}--\pageref{lastpage}}
\maketitle


\begin{abstract}
The redshift of $z\sim1.5$ is the transition epoch of protoclusters (PCs) from the star-forming phase into the quenching phase, and hence an appropriate era to investigate the build up of the quenched population. We define a `core' as the most massive halo in a given PC, where environmental effects are likely to work most effectively, and search for cores at $1<z<1.5$. We use a photometric redshift catalogue of a wide (effective area of $\sim22.2\,\mathrm{deg}^{2}$) and deep ($i\sim26.8\,\mathrm{mag}$) optical survey with Subaru Hyper-Suprime Cam. Regarding galaxies with $\log(M_{*}/M_{\odot})>11.3$ as the central galaxies of PC cores, we estimate their average halo mass by clustering analysis and find it to be $\log(M_\mathrm{h}/M_{\odot})\sim13.7$. An expected mass growth by the IllustrisTNG simulation and the observed overdensities around them suggest that the PC cores we find are progenitors of present-day clusters.  Classifying our galaxy sample into red and blue galaxies, we calculate the stellar mass function (SMF) and the red galaxy fraction. The SMFs in the PC cores are more-top heavy than field, implying early high-mass galaxy formation and disruption of low-mass galaxies. We also find that the red fraction increases with stellar mass, consistent with stellar-mass dependent environmental quenching recently found at $z>1$. Interestingly, although the cores with red and blue centrals have similar halo masses, only those with red centrals show a significant red fraction excess compared to the field, suggesting a conformity effect. Some observational features of PC cores may imply that the conformity is caused by assembly bias.
\end{abstract}




\section{Introduction}
Galaxies in the present-day Universe have great diversities in their morphology, colours, stellar mass, etc. These diversities are the result of numerous physical processes that have operated over cosmic time. In general, galaxy star formation activity depends on redshift as well as stellar mass and environment (e.g., \citealp{Dressler1980,Butcher1984,Cooper2007,Kodama2007,Peng2010,Wetzel2012,Darvish2016,Kawinwanichakij2017,Moutard2018,Lemaux2020,Chartab2020}). Therefore, quantifying galaxy properties as a function of redshift, stellar mass, and environment is important for understanding the physical drivers of galaxy formation and evolution.

The most massive structures in the present-day Universe are galaxy clusters, defined as virialized systems with a halo mass of $M_\mathrm{h}>10^{14}\,M_{\odot}$ (e.g. \citealp{Kravtsov_Borgani_2012,Overzier2016}). Clusters are the densest environments in the Universe, hosting hundreds to thousands of galaxies and hot plasma.
At $z<1$, galaxies in clusters are dominated by quiescent galaxies with elliptical morphologies and old stellar populations, largely different from those in the general field (e.g., \citealp{Dressler1980,Bower1998,Goto2003,Wetzel2012}). These quiescent galaxies are thought to be built up through many environmental effects in cluster host haloes: gas stripping by ram pressure and tidal force, interactions, harassment, starvation, etc. \citep{Gunn1972,Mihos2004,Moore1998,larson1980}. Moreover, galaxies can also be `pre-processed' when they are located in a group or a filament before entering cluster haloes (e.g., \citealp{DeLucia2012,Werner2022}). To reveal how these environmental effects, including pre-processing, have formed today's cluster galaxies, counterparts of clusters in the formation epoch are essential targets.

Because cluster formation takes a long cosmic time, most of the structure of a cluster progenitor has not been collapsed before $z=1$ \citep{Muldrew2015}. Before this epoch, galaxies that will belong to a cluster in the future are widely distributed in space, spanning several tens of comoving $\mathrm{Mpc}$ \citep{Chiang2013,Muldrew2015}. Such an extended structure is termed a protocluster (PC). As described below, the most massive halo in a given PC, or a `core'\footnote{Massive haloes (e.g., $M_\mathrm{h}\gtrsim10^{13-14}\,M_{\sun}$) at $z>1$ are sometimes called high-redshift groups or clusters. Since they will grow through the accretion from the surrounding regions until $z=0$, we regard such systems as PC cores at least at $z>1$ as \citet{Ando2020}.}, is an outstanding place, where accelerated galaxy formation and evolution occur \citep{Kodama2001,Finn2005,Koyama2010,shimakawa2014,Smail2014,Kato2016}.

On the simulation side, semi-analytic models applied to N-body simulations have been used to trace galaxy formation in PCs. \citet{Chiang2017} have shown that the star formation activity in PCs is divided into three stages: the `inside-out-growth' phase at $5\lesssim z\lesssim10$, the `extended star formation' phase at $1.5\lesssim z\lesssim5$, and the `infalling and quenching' phase at $z\lesssim1.5$. Before $z\sim 5$, active star formation occurs first in cores: at $z\sim10$, $\sim70$ per cent of all stars in PCs is formed in cores although they occupy only a small fraction of cosmic volume. Then, star formation gradually becomes more and more active in regions outside the cores. At $1.5\lesssim z \lesssim 5$, the star formation activity in PCs is at its peak with a total star formation rate (SFR) of $\sim1000\,M_{\odot}\,\mathrm{yr}^{-1}$. In some PC cores, a clear signal of galaxy quenching is seen near the end of this epoch. After $z\sim1.5$, the whole PC regions start to violently collapse into the cores, and galaxy quenching is accelerated.

\citet{Muldrew2018}, another simulation study, have found that the peak epoch of star formation is $\sim0.7\,\mathrm{Gyr}$ earlier in PCs than in the field, and quenching is enhanced since at least $z=3$, primarily in core regions. They have also reported a top-heavy stellar mass function (SMF) and a higher fraction of quiescent galaxies in PCs compared to the field, suggesting an enhancement of massive galaxy formation and accelerated quenching, especially in core regions.

On the observation side, enhanced star formation has been reported in cores as well as the rest of the PCs at $z\sim4$ \citep{Miller2018,Oteo2018,Ito2020} and at $z=2.5$ \citep{Wang2016,Shimakawa2018}. At $1.5<z<2.5$, PC cores dominated by the quiescent population have also been found \citep{Strazzullo2013,Newman2014,Cooke2016,Lee-Brown2017,Willis2020}. Below $z=1.5$, somewhat matured clusters or groups show a high fraction of quenched galaxies compared to the field (e.g., \citealp{Balogh2016,vanderBurg2018,vanderBurg2020,Reeves2021,Sarron2021}).

Both simulations and observations suggest that PC cores are places where galaxy formation and evolution are accelerated. In particular, PC cores at the transition epoch of star formation activity, $1<z<1.5$, are unique laboratories to investigate how environmental effects work to quench galaxy star formation in dense environment. In this paper, we focus on PC cores at $1<z<1.5$ and examine properties of galaxies within them.

One caveat in previous studies is that some observations are biased to progenitors of the most massive clusters ($M_\mathrm{h}>10^{15}\,M_{\odot}$ at $z=0$). Such massive clusters are very rare and only account for at most a few per cent of all clusters at $z=0$ (e.g., \citealp{Sheth1999,Chiang2013}). Another caveat is that the sample sizes of these studies are small. In fact, because individual clusters have very different assembly histories, the halo masses of cores have a large scatter ($\sim 1\,\mathrm{dex}$) at a fixed redshift and in descendant mass at $z=0$ \citep{Muldrew2015}. One may derive a biased picture of PC core galaxies if one only sees progenitors of most massive clusters or uses a small sample. To avoid this, one needs to construct a large PC core sample including not only massive ones but also less massive ones.

In \citet{Ando2020}, we have developed a new method to search for PC cores from a photometric redshift (photo-\textit{z}) catalogue at $1.5<z<3$. We have first selected possible tracers of PC cores and estimated their host halo masses. Then, we have confirmed that they grow into the cluster mass regime ($M_\mathrm{h}>10^{14}\,M_{\odot}$) by $z=0$. According to the stellar-to-halo mass relation (e.g., \citealp{Behroozi2013}), massive haloes with $M_{\mathrm{h}}\gtrsim 10^{13}\,M_{\sun}$ typically host high-mass central galaxies with $M_{*}\gtrsim 10^{11}\,M_{\sun}$, suggesting the possibility that such high-mass galaxies can be used as tracers of PC cores. However, there is a concern that high-mass galaxies are not always hosted by massive haloes with $M_{\mathrm{h}}\gtrsim 10^{13}\,M_{\sun}$ since the stellar-to-halo mass relation has a relatively shallower slope at $M_{\mathrm{h}}> 10^{12.5}\,M_{\sun}$. To isolate PC cores, multiple systems of massive galaxies can be used \citep{Diener2013,Bethermin2014,Ando2020}. \citet{Ando2020} have used pairs of massive galaxies as tracers and detected $75$ PC core candidates. However, identifying multiple galaxy systems is difficult when spectroscopic redshifts (spec-\textit{z}) are not available. Indeed, the massive galaxy pairs found in \citet{Ando2020} contain 46 per cent of contaminants by chance projection.

Another idea is to use extremely massive galaxies as tracers. \citet{Cheema2020} have shown that ultra-massive ($M_{*}\sim10^{11.5}\,M_{\odot}$) and passive galaxies at $z\sim1.6$ are hosted by haloes more massive than $1\times 10^{14}\,M_{\odot}$, and they have no massive neighbours with identical stellar masses \citep{Sawicki2020}. Based on these facts, we use single massive galaxies as tracers of PC cores. Unlike \citet{Cheema2020}, we use both star-forming and quiescent galaxies in our search to investigate whether the star formation activities of satellite galaxies differ depending on those of central galaxies, e.g. galactic conformity \citep{Weinmann2006}.

To construct a statistical sample of PC cores, we need a large survey volume since PC cores are very rare objects with number density of $\sim10^{-5}\,\mathrm{cMpc}^{-3}$. Besides, deep photometric data are required to examine the properties of satellite galaxies in PC cores since they are much fainter than massive central galaxies. In this study, we use data from an optical imaging survey named the Hyper-Suprime Cam~(HSC; \citealp{Miyazaki2018,Komiyama2018,Furusawa2018}) Subaru Strategic Program~(HSC-SSP; \citealp{Aihara2018a, Aihara2021}). The HSC-SSP provides a very deep and wide data set ($\sim30\,\mathrm{deg}^{2}$, $i\sim 27\text{--}28\,\mathrm{mag}$), which is suitable for a PC core search.

In this paper, with a statistical sample of PC cores, we calculate the SMF and the red galaxy fraction, which reflect the cumulative (past) and the differential (current) star formation activity, respectively. Focusing on these two statistical quantities, we examine how quenching in PC cores progresses at the transition epoch of their star formation activity.

The structure of this paper is as follows. In \S~2, we describe the data and galaxy samples used in this study. In \S~3, we introduce the method to find PC cores and describe the results of our search. We validate the obtained PC core candidates from comparisons with the IllustisTNG simulation and observed galaxy overdensity profiles. In \S~4, we examine properties of member galaxies in the PC core candidates focusing on the SMF and the red galaxy fraction. In \S~5, we compare our results with the literature and discuss the formation of PC core galaxies. \S~6 is devoted to a summary and conclusions.
 
Throughout this paper, we assume a flat $\mathrm{\Lambda}$CDM cosmology with $(\Omega_\mathrm{m},\, \Omega_\mathrm{\Lambda},\, h,\, \sigma_\mathrm{8},\, n_{0})=(0.3,\, 0.7,\, 0.7, \,0.81,\, 0.95)$ and a \citet{Chabrier2003} initial mass function. We use AB magnitudes \citep{Oke1983} and the notations $\mathrm{cMpc}$ and $\mathrm{pMpc}$ to indicate comoving and physical scales, respectively.

\section{Data and sample selection}

\subsection{Photometric data}
We use data from the Public Data Release 3 (PDR3) of the HSC-SSP. The HSC-SSP is a very wide and deep optical imaging survey with five broad bands (\textit{grizy}, \citealp{Kawanomoto2018}), including three layers: Wide ($1400\,\mathrm{deg}^{2}$, $i\sim 26\,\mathrm{mag}$), Deep (D, $26\,\mathrm{deg}^{2}$, $i\sim 27\,\mathrm{mag}$) and UltraDeep (UD, $3.5\,\mathrm{deg}^{2}$, $i\sim 28\,\mathrm{mag}$). The D layer is composed of four separated fields called E-COSMOS, XMM-LSS, ELAIS-N1, and DEEP2-3. The two UD regions, called COSMOS and SXDS, are adjacent to E-COSMOS and XMM-LSS. In this study, we use data taken in the D and UD (DUD) layers to derive reliable photometry and thus a photo-\textit{z} even for faint sources. We only use areas within $0.75\,\mathrm{deg}$ from the fiducial pointings of each DUD field to exclude data that are too shallow.

The HSC-SSP PDR3 data are processed with \texttt{hscPipe 8} \citep{Bosch2019}. The \texttt{hscPipe} is based on software developed for the pipeline of the Large Synoptic Survey Telescope (LSST; \citealp{Juric2017,Ivezic2019}). Masks for bad pixels and artefacts around bright stars are provided.

\begin{figure}
	\includegraphics[width=\columnwidth]{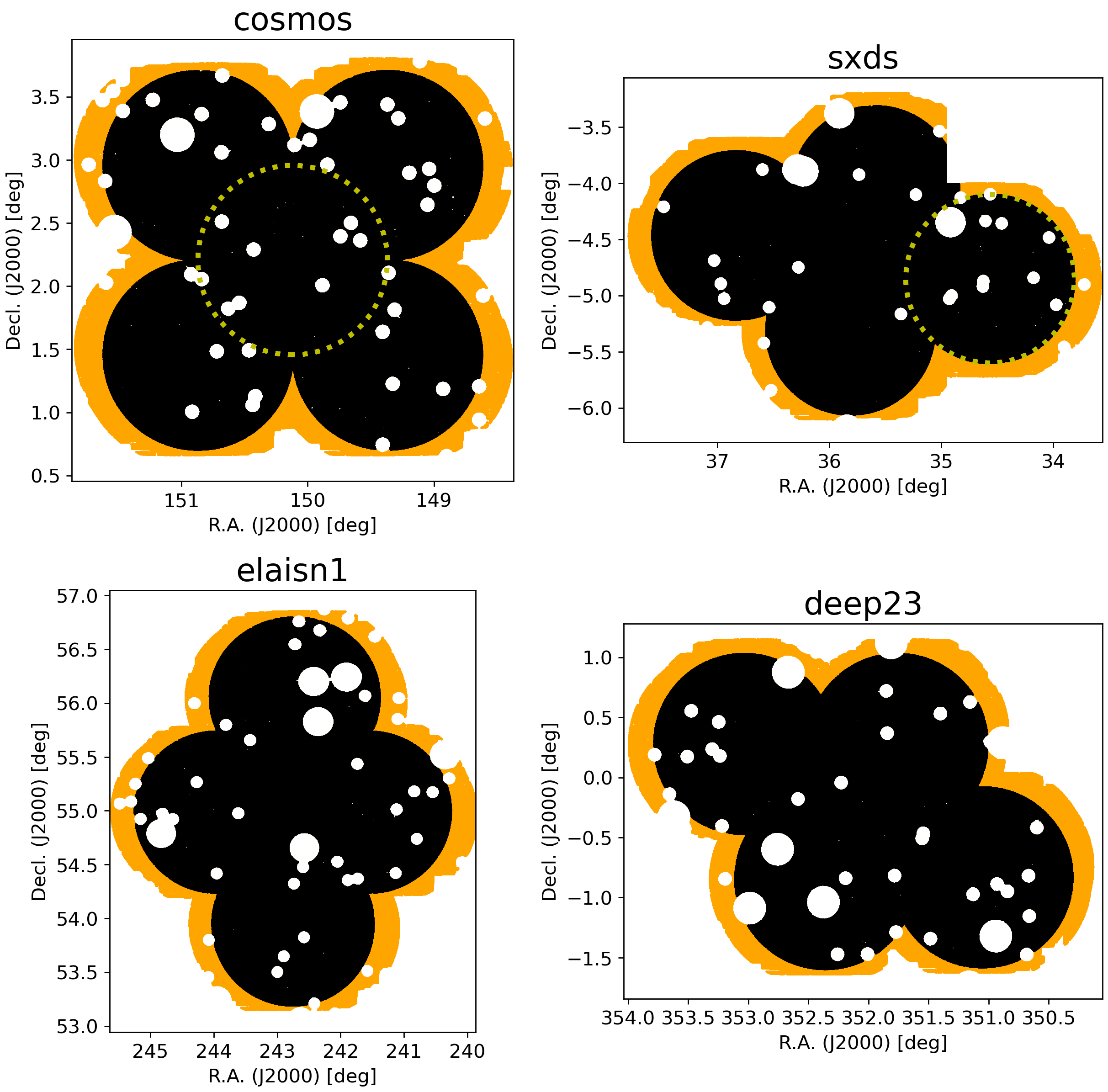}
    \caption{The survey footprints of the four DUD fields: COSMOS (top left), SXDS (top right), ELAIS-N1 (bottom left), and DEEP2-3 (bottom right). Areas with bad pixel and bright star masks are removed. We only use data within $0.75\,\mathrm{deg}$ from fiducial pointings (black areas) since data outside of these regions (orange areas) are not reliable (i.e., shallower data). Yellow dotted circles show the UD regions.}
    \label{fig:footprint}
\end{figure}

We set the following flags to \texttt{False} to remove bad pixels: \texttt{X\_pixelflags\_offimage},
\texttt{X\_pixelflags\_edge}, \texttt{X\_pixelflags\_saturatedcenter}, and \texttt{X\_pixelflags\_bad}, where X represents five bands (\textit{grizy}). We also set the following flags to False to avoid artefacts around bright stars: \texttt{X\_mask\_brightstar\_halo}, \texttt{X\_mask\_brightstar\_ghost}, \texttt{X\_mask\_brightstar\_blooming}, \texttt{X\_mask\_brightstar\_ghost15}, and \texttt{y\_mask\_brightstar\_scratch}. In addition, \texttt{X\_cmodel\_flag==False} is also applied to select objects with secure CModel magnitudes \citep{Abazajian2004,Bosch2018}. The total survey areas are summarised in Table \ref{tab:depth}. The survey footprints with these masks are shown in Fig.~\ref{fig:footprint}.

First, we select extended objects (i.e., galaxies) adopting \texttt{i\_extendedness\_value==1}. Then, we apply a magnitude cut since clustering analysis requires uniform source detection over the entire survey field. Table~\ref{tab:depth} shows the median values of $5\sigma$ limiting magnitudes for point sources\footnote{The UD regions are about one magnitude deeper than the D regions. However, each of the two UD regions occupies only a quarter of the corresponding D+UD area (see Fig.~\ref{fig:footprint}), the difference in observational depth between D and UD has only a small effect on the median depth.}. To select sources uniformly among the four DUD fields, we set the shallowest values to common magnitude limits for each of the \textit{r, i, z}, and \textit{y} bands: $r\leq27.1$, $i\leq26.8$, $z\leq26.4$, $y\leq25.2$. For the \textit{g} band, we apply no magnitude cut to include galaxies with very red colours in our analysis. We note that the results of this paper are not changed even if we also apply a \textit{g}-band magnitude cut.

\begin{table*}
\centering
    \caption{The median $5\sigma$ limiting magnitudes and the survey areas in the four DUD fields.}
    \label{tab:depth}
	\begin{tabular}{lccccccc}
		\hline
		field & $g$ & $r$ & $i$ & $z$ & $y$ & area & volume$^a$ \\
		   & mag & mag & mag & mag & mag & $\mathrm{deg}^2$ & $10^{7}\,\mathrm{cMpc}^{3}$ \\
		\hline
		COSMOS$^b$ &  $27.7$ & $27.3$ & $27.3$ & $26.8$ & $25.6$ & 6.21 & 3.00  \\  [2pt]
		SXDS$^c$ & $27.7$ & $27.2$ & $26.8$ & $26.4$ & $25.2$ & 5.61 & 2.71  \\  [2pt]
		ELAIS-N1 & $27.8$ & $27.1$ & $27.1$ & $26.4$ & $25.2$ & 5.07 & 2.45  \\  [2pt]
		DEEP2-3 & $27.3$ & $27.1$ & $26.8$ & $26.4$ & $25.5$ & 5.31 & 2.57  \\  [2pt]
		\hline
        magnitude cut & - & $27.1$ & $26.8$ & $26.4$ & $25.2$ & - & - \\  [2pt]
        total area & - & - & - & - & - & 22.2 & 10.7 \\  [2pt]
		\hline
	\end{tabular}
	\begin{tablenotes}[normal]
	 \item \textit{Notes.} $^a$$1<z<1.5$ is assumed. $^b$Deep E-COSMOS and UltraDeep COSMOS. $^c$Deep XMM-LSS and UltraDeep SXDS.
    \end{tablenotes}
\end{table*}

\subsection{Photometric redshift catalogue}
The HSC-SSP provides several photo-\textit{z} catalogues\footnote{The photo-\textit{z} catalogues of the HSC-SSP PDR3 are only internally available as of this writing. They will be open to the community in the near future.} \citep{Tanaka2018,Nishizawa2020}. We use a photo-\textit{z} catalogue based on a SED fitting code with Bayesian priors on physical properties of galaxies, called \textsc{MIZUKI} \citep{Tanaka2015}. \textsc{MIZUKI} uses SED templates from the \citet{Bruzual&Charlot2003} stellar population synthesis model, assuming a \citet{Chabrier2003} initial mass function, \cite{Calzetti2000}'s dust attenuation curve, exponentially decaying SFRs and Solar metallicity. See \citet{Tanaka2015,Tanaka2018,Nishizawa2020} for detailed information.

\begin{figure*}
	\includegraphics[width=2\columnwidth]{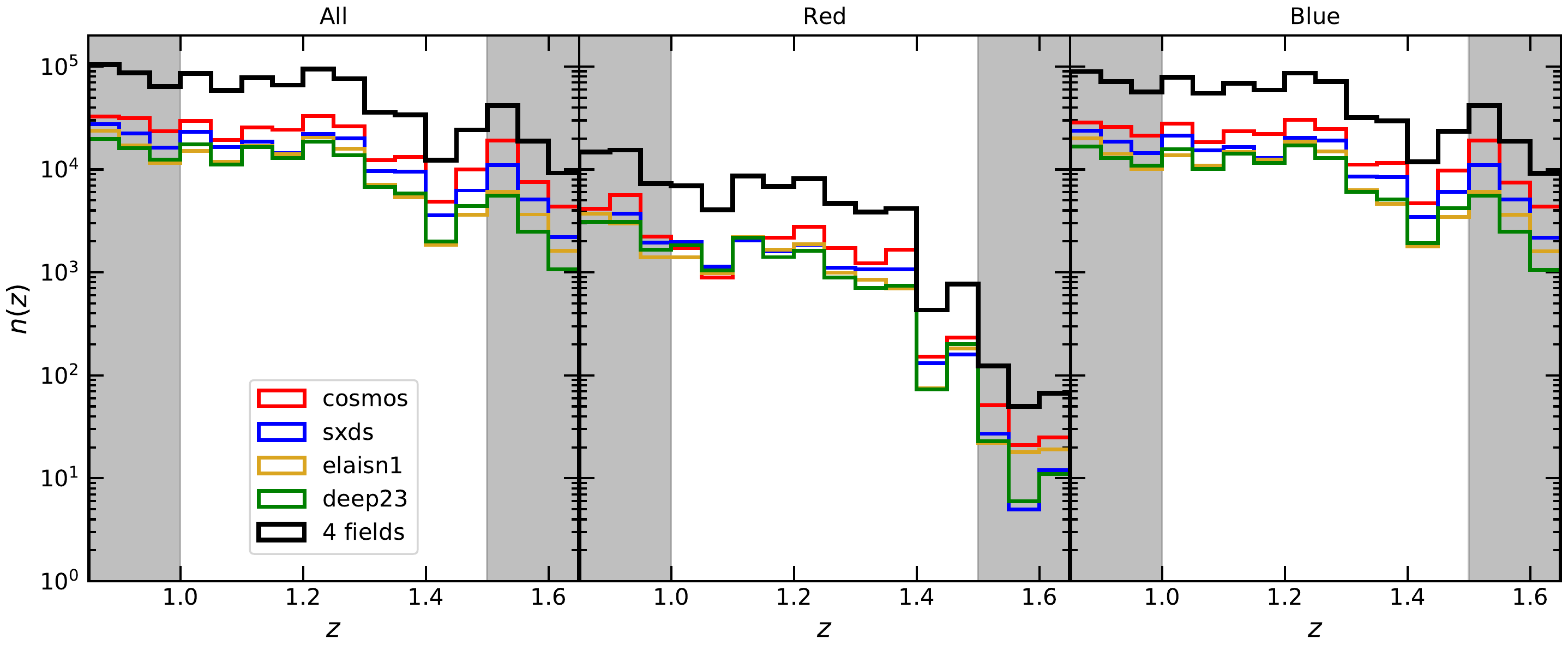}
    \caption{The redshift distributions of the main sample. The left, middle and right panels correspond to all, red, and blue galaxies, respectively. Red, blue, yellow, green, and black lines indicate COSMOS, SXDS, ELAIS-N1, DEEP2-3, and the sum of the four fields. We do not focus on $z<1$ and $z>1.5$ (grey shaded areas) in the main analysis except for overdensity estimation (\S~\ref{sec:overdensity}) and characterisation of field galaxies (\S~\ref{sec:properties}).}
    \label{fig:n_z}
\end{figure*}

\begin{figure*}
	\includegraphics[width=2\columnwidth]{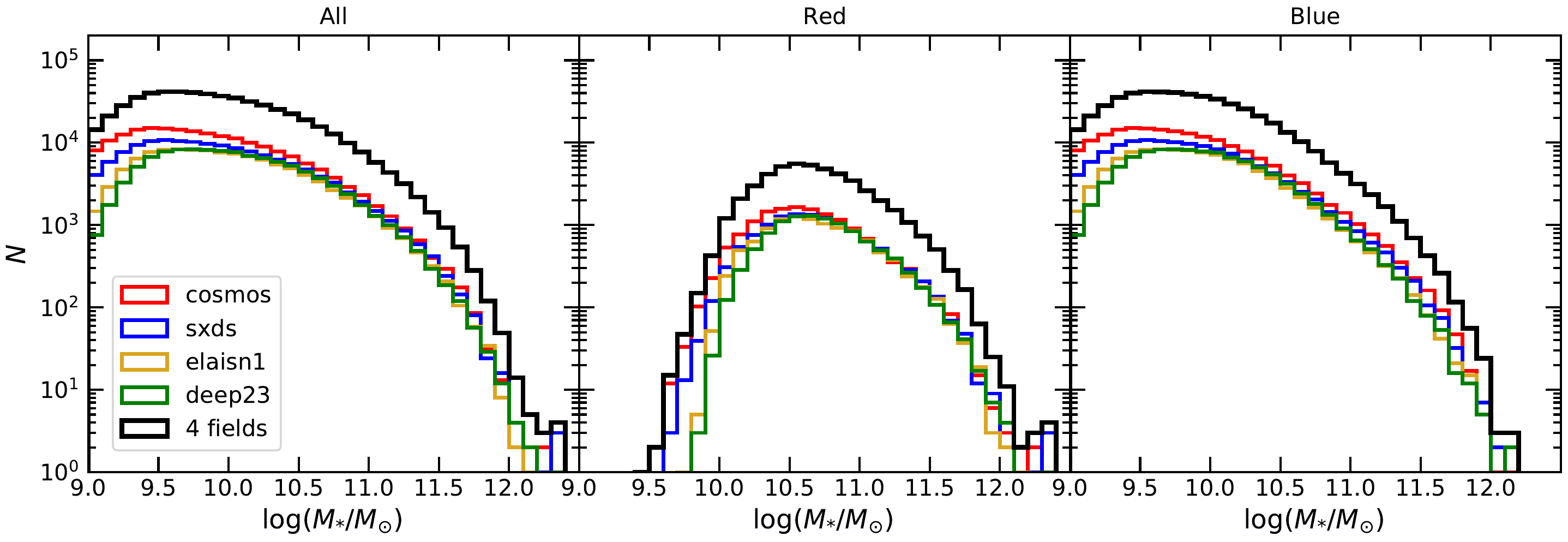}
    \caption{The stellar mass distributions of the main sample at $1<z<1.5$. The meaning of the symbols is the same as Fig.~\ref{fig:n_z}.}
    \label{fig:n_mass}
\end{figure*}

To ensure reliable photo-\textit{z}'s, we only use objects with \mbox{$\chi^{2}_{\nu}<3$} and \mbox{\texttt{risk}$<0.1$}, where $\chi^{2}_{\nu}$ is the reduced chi-squares of the best-fit model and \texttt{risk} is an indicator of the probability that the estimated photo-\textit{z} is an outlier. We also require \mbox{\texttt{prob\_gal}$>0.9$}, where \texttt{prob\_gal} is the probability that the object is a galaxy. Then, we select all objects with $0.85 \leq z \leq 1.65$ and $\log(M_{*}/M_{\odot})\geq 9$. The left panels of Fig.~\ref{fig:n_z} and Fig.~\ref{fig:n_mass} show the redshift and stellar mass distributions of galaxies in the four DUD fields. Since the scope of this study is PC cores at $1<z<1.5$, we only show the stellar mass distribution of galaxies at $1<z<1.5$ in Fig.~\ref{fig:n_mass}. Galaxies at $z<1$ and $z>1.5$ are used for overdensity estimation (\S~\ref{sec:overdensity}) and characterisation of field galaxies (\S~\ref{sec:properties}).

In addition to photo-\textit{z}'s, the HSC-SSP provides spec-\textit{z}'s for some fraction of galaxies gathered from various studies (see \S~4.1 of \citealp{Aihara2021}). To test the photo-\textit{z} precision, we calculate the bias $b_{z}$ and normalised median absolute deviation $\sigma_{z}$ as:
\begin{flalign}
\label{sigma_nmad}
        b_{z} &= \mathrm{median}\left( \Delta z \right),  \\
        \sigma_{z} &= 1.48\times \mathrm{median}\left( \left| \frac{|\Delta z-\mathrm{median}(\Delta z)|}{1+z_\mathrm{spec}}\right| \right),  \\
        \Delta z &= z_\mathrm{phot}-z_\mathrm{spec},
\end{flalign}
where $z_\mathrm{phot}$ and $z_\mathrm{spec}$ are photo-\textit{z} and spec-\textit{z}, respectively. The normalised median absolute deviation is similar to the normal standard deviation but less sensitive to outliers \citep{Brammer2008}. We also define outliers as galaxies with $|\Delta z| >0.15(1+z_\mathrm{spec})$. Here, we only use galaxies at $1<z<1.5$. We summarise the numbers of galaxies with spec-\textit{z}, $\sigma_{z}$, $b_{z}$, and the fraction of outliner galaxies, $\eta$, in Table~\ref{tab:photoz_prec}. In addition, we present the photo-\textit{z} precision as a function of redshift in Appendix~\ref{app:photoz_prec}.

It has been argued that the stellar masses estimated by \textsc{MIZUKI} with the HSC-SSP data are slightly ($\sim0.2\,\mathrm{dex}$) higher than those obtained by some other SED fitting codes and multi-wavelength data\footnote{A stellar mass offset is not unique to the HSC-SSP and \textsc{MIZUKI} but commonly seen even among galaxy catalogues with multi-wavelength data including near-infrared photometry (e.g., \citealp{vanDokkum2014}).} \citep{Tanaka2015,Tanaka2018}. We compare the stellar masses of our sample with those in the COSMOS2015 catalogue (\citealp{Laigle2016}; L16 hereafter) and examine how different stellar mass estimates affect the results of this paper in Appendix~\ref{app:robustness}. In brief, the main results are robust against uncertainties in stellar mass estimates.

\begin{table*}
\centering
    \caption{The photo-\textit{z} precision of the main sample and the cross-matched sample with COSMOS2015.}
    \label{tab:photoz_prec}
	\begin{tabular}{cccccccccccccccc}
		\hline
		& \multicolumn{4}{c}{all} & $\ $ & \multicolumn{4}{c}{red} & $\ $ & \multicolumn{4}{c}{blue}  \\
		\cline{2-5} \cline{7-10} \cline{12-15}
		mass & N & $\sigma_{z}$& $b_{z}$ & $\eta$ & & N & $\sigma_{z}$ & $b_{z}$ & $\eta$ & & N &$\sigma_{z}$& $b_{z}$ & $\eta$ \\
		$\log(M_{*}/M_{\odot})$ & \# & & & \% & & \# & & & \% & & \# & & & \% \\
		\hline
		$[9,10) $ & $5959$ & $0.031$ & $-0.024$ & $6.3$ & & $25$ & $0.173$ & $-0.114$ & $44.0$ & & $5934$ & $0.031$ & $-0.024$ & $6.1$ \\  [2pt]
        $[10,11) $ & $8480$ & $0.033$ & $-0.005$ & $4.3$ & & $1008$ & $0.052$ & $0.002$ & $3.7$ & & $7472$ & $0.033$ & $-0.006$ & $4.4$ \\  [2pt]
        $[11,\inf) $ & $1687$ & $0.037$ & $-0.003$ & $9.8$ & & $713$ & $0.030$ & $-0.031$ & $3.1$ & & $974$ & $0.037$ & $0.019$ & $14.7$ \\  [2pt]
		\hline \hline
		& \multicolumn{4}{c}{all} & $\ $ & \multicolumn{4}{c}{QG} & $\ $ & \multicolumn{4}{c}{SFG}  \\
		\cline{2-5} \cline{7-10} \cline{12-15}
		mass & N & $\sigma_{z}$& $b_{z}$ & $\eta$ & & N & $\sigma_{z}$ & $b_{z}$ & $\eta$ & & N &$\sigma_{z}$& $b_{z}$ & $\eta$ \\
		$\log(M_{*}/M_{\odot})$ & \# & & & \% & & \# & & & \% & & \# & & & \% \\
		\hline
        $[9,10) $ & $1912$ & $0.027$ & $-0.016$ & $2.3$ & & $6$ & $0.078$ & $-0.097$ & $16.7$ & & $1906$ & $0.027$ & $-0.016$ & $2.3$ \\  [2pt]
        $[10,11) $ & $2122$ & $0.035$ & $-0.001$ & $2.7$ & & $263$ & $0.051$ & $0.004$ & $2.3$ & & $1859$ & $0.035$ & $-0.001$ & $2.8$ \\  [2pt]
        $[11,\inf) $ & $441$ & $0.039$ & $0.006$ & $11.3$ & & $179$ & $0.033$ & $-0.032$ & $3.4$ & & $262$ & $0.039$ & $0.027$ & $16.8$ \\  [2pt]
		\hline
	\end{tabular}
	\begin{tablenotes}[normal]
	\item \textit{Notes.} The upper side shows our main sample with our RG/BG classification, while the lower side shows the cross-matched subsample with COSMOS2015 with QG/SFG classification by \citet{Laigle2016}.
    \end{tablenotes}
\end{table*}

\subsection{Galaxy classification}
\label{sec:class}
Galaxy colour can be used as an indicator of star formation: red and blue colours indicate quenched and active star formation, respectively (e.g., \citealp{Salim2014}). To investigate the environmental dependence of star formation activity, we define star formation class using galaxy colours.

At $1<z<1.5$, the HSC \textit{grizy} bands correspond to rest-frame UV to $\sim5000$~\AA. We use the rest-frame $\mathrm{NUV}-g$ colour derived from SED fitting to classify galaxies, where \textrm{NUV} is the magnitude in the GALEX \textrm{NUV} band \citep{Martin2005}. We define galaxies with $M_\mathrm{NUV}-M_{g} \geq 3.2$ as red galaxies (RGs) and the others as blue galaxies (BGs). The boundary is determined as described later. The redshift and stellar mass distributions for RGs and BGs are shown in the middle and right panels of Fig.~\ref{fig:n_z} and Fig.~\ref{fig:n_mass}. In general, a single colour criterion cannot distinguish quiescent galaxies (QGs) from dust-rich star-forming galaxies (SFGs), which typically have red colours. To evaluate how completely and purely this criterion can select QGs and SFGs, we compare our classification results with those provided in the COSMOS2015 catalogue (L16), whose footprint largely overlaps with the UD COSMOS.

The COSMOS2015 catalogue contains physical quantities such as photo-\textit{z}, rest-frame colours, stellar mass, etc., derived from SED fitting performed by \texttt{LAPHARE} code \citep{Arnouts2002,Ilbert2006} with about 30 band photometry from \textrm{NUV} to near infrared. They use the $\mathrm{NUV}-r$ versus $r-J$ colour-colour plane to classify galaxies: QGs are defined as those satisfying $M_\mathrm{NUV}-M_{r}>3(M_{r}-M_{J})+1$ and $M_\mathrm{NUV}-M_{r}>3.1$, and SFGs are the others. This criterion successfully distinguishes QGs from dusty SFGs \citep{Williams2009,Ilbert2013}.

First, we cross-match galaxies in our sample and COSMOS2015 within $1\arcsec$ separation. Then, we calculate selection completeness and purity as following:
\begin{flalign}
    \label{comp}
    \mathrm{select.\ comp.}&=
    \begin{cases}
        \frac{N(\mathrm{RG|QG})}{N(\mathrm{RG|QG})+N(\mathrm{BG|QG})} & \mathrm{(for\ RG/QG)}  \\[6pt]
        \frac{N(\mathrm{BG|SFG})}{N(\mathrm{RG|SFG})+N(\mathrm{BG|SFG})} & \mathrm{(for\ BG/SFG)},
    \end{cases}\\[6pt]
    \label{purity}    
    \mathrm{purity}&=
    \begin{cases}
        \frac{N(\mathrm{RG|QG})}{N(\mathrm{RG|QG})+N(\mathrm{RG|SFG})} & \mathrm{(for\ RG/QG)}  \\[6pt]
        \frac{N(\mathrm{BG|SFG})}{N(\mathrm{BG|QG})+N(\mathrm{BG|SFG})} & \mathrm{(for\ BG/SFG)},
    \end{cases}
\end{flalign}
where $N(\mathrm{X|Y})$ is the number of galaxies classified as $\mathrm{X=\{RG,BG\}}$ and $\mathrm{Y=\{QG,SFG\}}$. We calculate selection completeness and purity for RG/QG using different $M_\mathrm{NUV}-M_\mathrm{g}$ boundary values from $0$ to $6$ to explore the trade-off between completeness and purity. We find that both selection completeness and purity for QGs are about 80 per cent with a boundary value of $3.2$. As a general trend, they are higher for higher mass galaxies: at $1<z<1.5$, selection completeness and purity are 82 per cent and 79 per cent at $\log(M_{*}/M_{\odot})\geq10$, respectively, while 31 per cent and 57 per cent at $\log(M_{*}/M_{\odot})\leq10$, respectively. In terms of redshift, selection completeness is worse for higher redshift galaxies: at $\log(M_{*}/M_{\odot})\geq10$, selection completeness is 83 per cent at $1<z<1.4$, while 54 per cent at $1.4<z<1.5$. In this sense, our classification might not be valid for galaxies with $\log(M_{*}/M_{\odot})\leq10$, and it is somewhat uncertain at $z>1.4$. In addition to RG/QG, we also calculate selection completeness and purity for BG/SFG adopting the boundary value of $3.2$ and find both of them to be higher than roughly 95 per cent at any mass and redshift ranges.

\begin{figure}
	\includegraphics[width=\columnwidth]{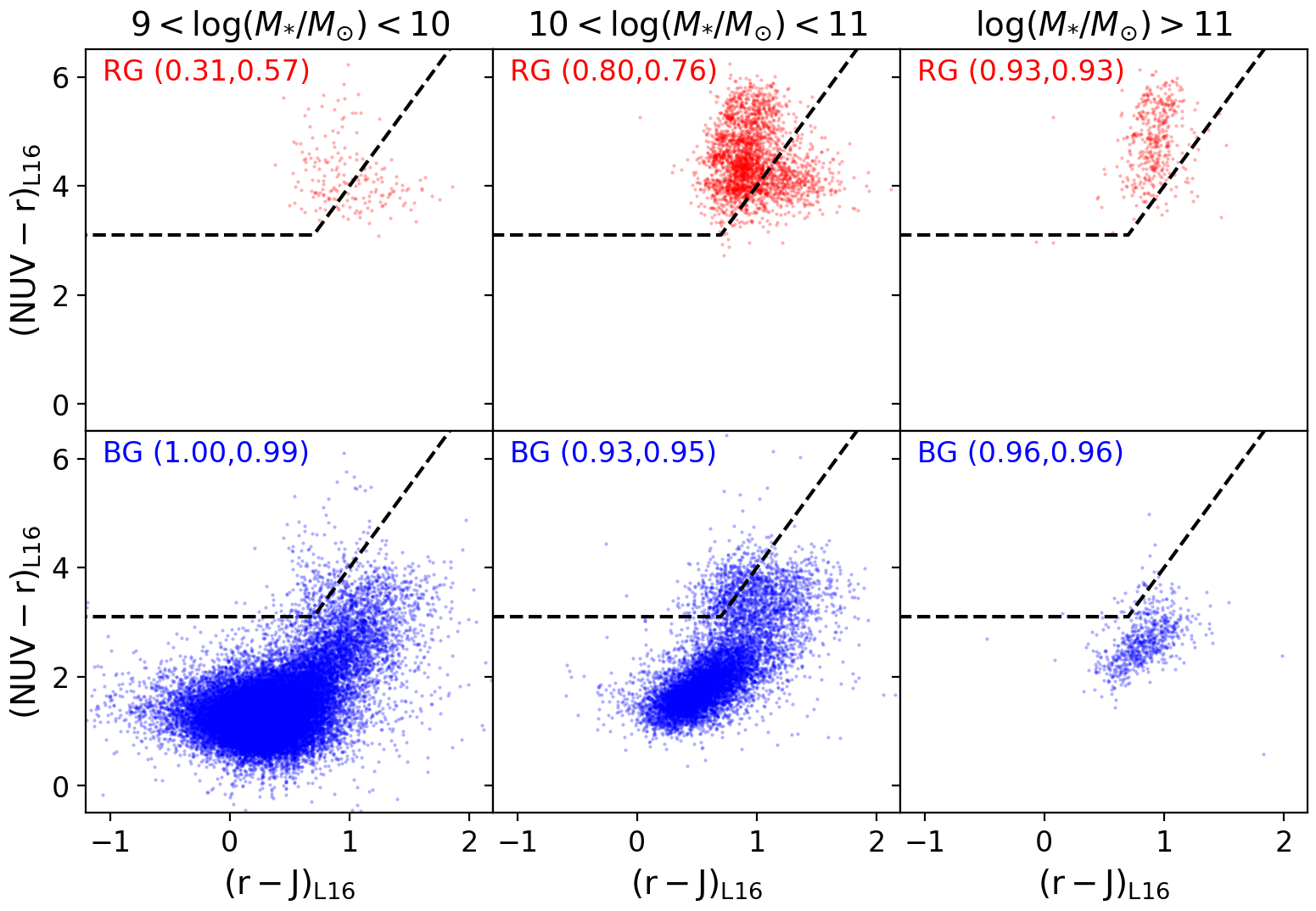}
    \caption{The distributions of RGs (top) and BGs (bottom) at $1<z<1.5$ on the two rest-frame colour plane ($\mathrm{NUV}-r$ versus $r-J$) measured by \citet{Laigle2016} (L16). Each column shows a different stellar mass range. Galaxies located in the upper left region enclosed by the dashed line are classified as QGs. The bracket at the top left corner of each panel shows selection completeness (left) and purity (right).}
    \label{fig:class_qg_sfg}
\end{figure}

We visually compare our classification and that of L16 in Fig.~\ref{fig:class_qg_sfg}. We separately plot RGs (top) and BGs (bottom) at $1<z<1.5$ on the rest-frame $\mathrm{NUV}-r$ and $r-J$ colour-colour plane measured by L16. Galaxies located in the upper left region enclosed by the dashed line are classified as QGs. We also show selection completeness (left) and purity (right) in the bracket at the top left corner of each panel. This figure shows that our single colour classification works well at least for $\log(M_{*}/M_{\odot})>10$.

We note that the galaxy classification in COSMOS2015 also contains some uncertainties (e.g., uncertainties in rest-frame colour estimates). We also note that some fraction of `contaminants' for RGs might be green valley galaxies (i.e., transitioning galaxy from SFG to QG) rather than dusty SFGs.

\subsection{Stellar mass completeness}
We estimate the stellar mass completeness of our sample following an empirical method (e.g., \citealp{Pozzetti2010, Davidzon2013, Ilbert2013}; L16). First, for each galaxy in the sample, we calculate the re-scaled stellar mass, $M_{*,\mathrm{res}}$, that a galaxy at the same redshift but at the limiting magnitude will have:
\begin{equation}
\label{m_lim}
    M_{*,\mathrm{res}}=\log(M_{*})-0.4(X_{\mathrm{lim}}-X),
\end{equation}
where $X$ and $X_{\mathrm{lim}}$ are the observed and limiting magnitudes in band X, respectively. Using \textit{i}-band magnitudes, we calculate $M_{*,\mathrm{res}}$ separately for RGs and BGs with a redshift interval of $\Delta z=0.1$. At a fixed redshift, the detection completeness at a given stellar mass $M$ is defined as the fraction of the $M_{*,\mathrm{res}}$ distribution below $M$:

\begin{equation}
\label{mass_comp}
    \mathrm{detect.\ comp.}(z_{i},M)=\frac{N(z_{i},M_{*,\mathrm{res}}<M)}{N(z_{i})},
\end{equation}
where $N(z_{i},M_{*,\mathrm{res}}<M)$ is the number of galaxies at \textit{i}-th redshift bin with smaller $M_{*,\mathrm{res}}$ than $M$, while $N(z)$ is the total number of galaxies at that redshift bin. 

\begin{figure}
	\includegraphics[width=\columnwidth]{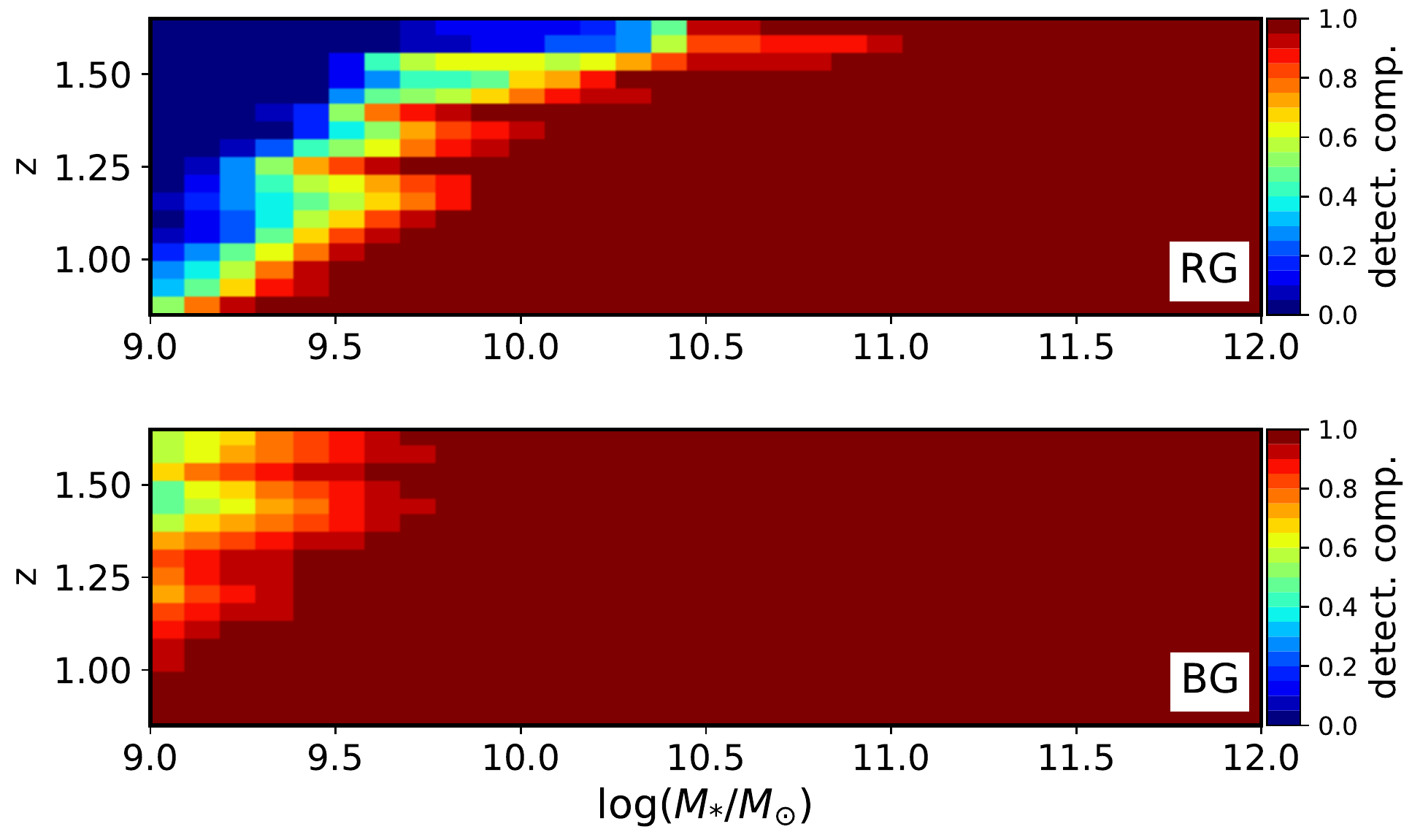}
    \caption{The derived detection completeness for the DUD COSMOS field. The detection completenesses for RGs and BGs are separately calculated and shown in the top and bottom panels, respectively. Most galaxies at $\log(M_{*}/M_{\odot})\geq10$ are detected. At $\log(M_{*}/M_{\odot})\leq10$, galaxies at higher redshift are not completely detected especially for RGs.}
    \label{fig:comp_m_z}
\end{figure}

Fig.~\ref{fig:comp_m_z} shows derived detection completeness on the redshift versus stellar mass plane for the DUD COSMOS field. There is no significant difference among the four DUD fields. Galaxies at $\log(M_{*}/M_{\odot})\geq10$ are almost completely detected at $1<z<1.5$. At $\log(M_{*}/M_{\odot})\leq10$, some fraction of galaxies at higher redshift is missed, especially for RGs. In the following sections of this paper, we mainly focus on the stellar mass range of $\log(M_{*}/M_{\odot})\geq10$, and we do not correct detection completeness when we calculate statistical quantities such as the SMF and the red fraction to avoid uncertainties due to large completeness correction.

\begin{figure}
	\includegraphics[width=\columnwidth]{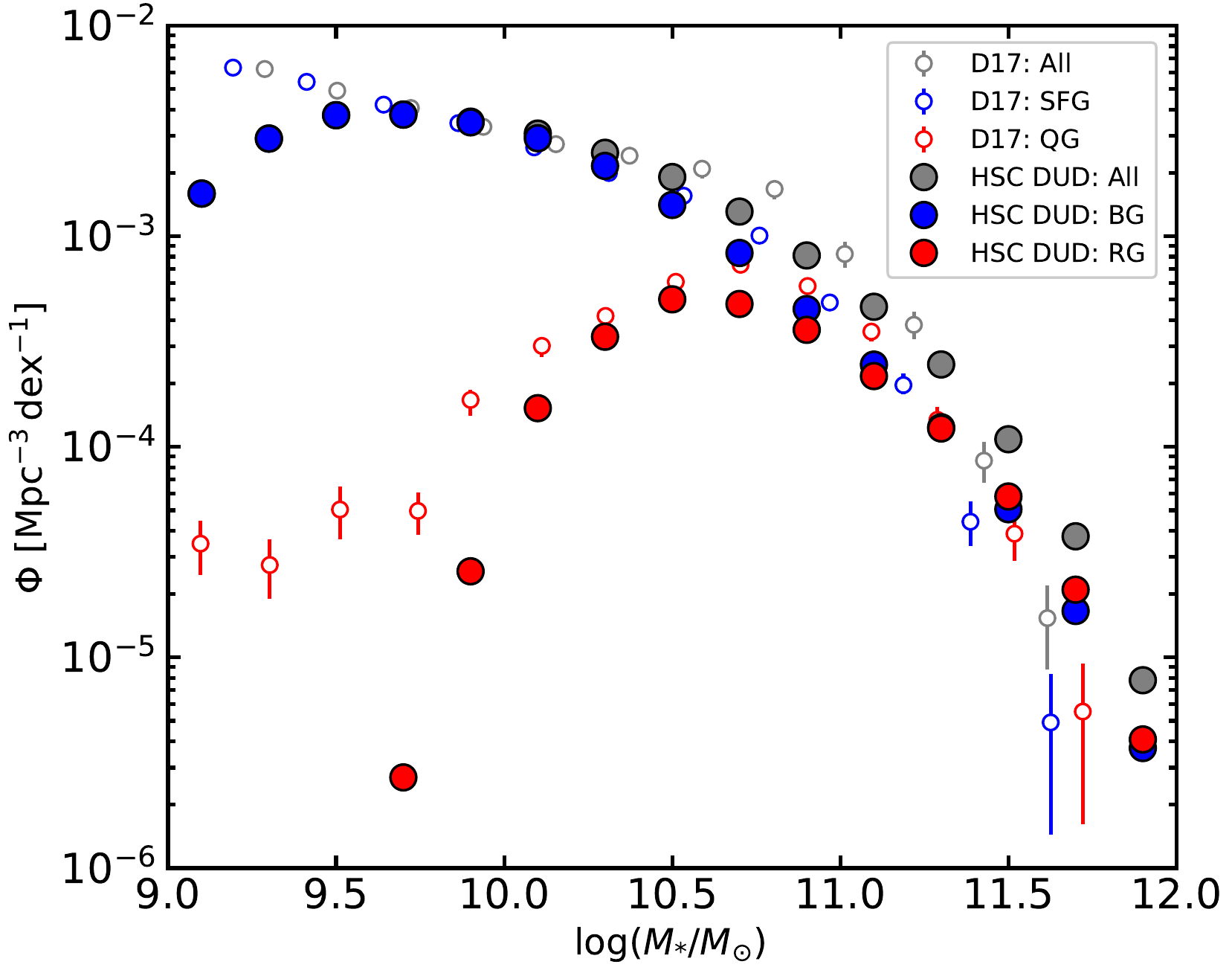}
    \caption{The SMFs of the HSC-SSP DUD galaxies and those of \citet{Davidzon2017} (D17). Filled and open circles show the data points of the HSC-SSP and D17, respectively. Gray, blue, and red colours mean the SMFs for all galaxies, BGs (SFGs) and RGs (QGs) of the HSC-SSP (D17). The SMFs of the HSC-SSP are calculated using galaxies at $1<z<1.5$, while those of D17 at $1<z<1.4$}
    \label{fig:comp_smf}
\end{figure}

Lastly, we compare the SMFs from our sample with those based on the COSMOS2015 catalogue (\citealp{Davidzon2017}; D17) in Fig.~\ref{fig:comp_smf}. We derive SMFs by dividing the number of galaxies at a given mass bin by our survey volume. D17's SMFs are for $1<z<1.4$ and have been corrected for completeness (their Fig.~15 and Fig.~16).

Our SMFs for RGs and BGs agree well with those of D17 at $10.3\leq\log(M_{*}/M_{\odot})\leq11.5$ and $9.5\leq \log(M_{*}/M_{\odot})\leq11.5$, respectively, and are lower than D17's at $\log(M_{*}/M_{\odot})\lesssim10.3$ and $\log(M_{*}/M_{\odot})\lesssim9.5$, respectively, being consistent with our mass completeness limit. At a higher-mass range ($\log(M_{*}/M_{\odot})\geq11.5$), our SMFs are larger by several times than those of D17. To examine the cause of this, we compare the stellar masses of \textsc{MIZUKI} with those of COSMOS2015. We find that at this high mass range, \textsc{MIZUKI} has a relatively large scatter if the mass of COSMOS2015 is fixed. Qualitatively, this scatter can contribute to an apparent overabundance of high-mass galaxies through the Eddington bias. We do not discuss this topic further since a detailed comparison of the stellar mass estimates is out of the scope of this study. In any case, we conclude that our galaxy selection and classification are plausible.

\section{Construction of a protocluster core sample}
PC cores are massive haloes which grow into $\log(M_\mathrm{h}/M_{\odot})\gtrsim 14$ clusters by $z=0$. Their typical mass at $1<z<1.5$ is $\log(M_\mathrm{h}/M_{\odot})\gtrsim 13.5$ \citep{Behroozi2013}. According to the abundance matching technique, the typical stellar mass of central galaxies hosted by such massive halos is $\log(M_{*}/M_{\odot})>11$ (e.g., \citealp{Behroozi2013}).

Considering these facts, we search for PC cores as follows. First, we select galaxies more massive than a range of mass limit, $10.8\leq \log(M_{*,\mathrm{lim}}/M_{\odot})\leq 11.5$, at $1<z<1.5$. They are regarded as candidate central galaxies of PC cores. Next, we estimate their host halo mass by clustering analysis. Then, we examine the halo mass evolution of the PC core candidates using mock halo catalogues of the IllustrisTNG \citep{Pillepich2018a}, cosmological hydrodynamical simulations, to see whether they grow into the cluster mass regime, $\log(M_\mathrm{h}/M_{\odot})\gtrsim14$ by $z=0$. As an additional test, we also calculate overdensity profiles around central galaxies to check the detected PC core candidates are actually located in overdense regions.

\subsection{Clustering analysis}
\label{sec:clustering}
We estimate halo masses using almost the same way as \citet{Ando2020} and briefly describe the method below.

We first calculate the two-point angular auto-correlation function (ACF) of the selected massive galaxies $\omega(\theta)$ and its errors $\varepsilon_{\omega}(\theta)$ assuming Poisson errors\footnote{With a relatively large sample ($\sim10^{5\text{--}6}$), the Poisson errors may underestimate the uncertainties in the ACF, while other resampling approaches such as bootstrapping and jackknifing may overestimate them \citep{Norberg2009,Khostovan2018}. However, the Poisson errors are consistent with those estimated by a bootstrapping when the sample size is small enough, at least up to an order of $10^{3}$ \citep{Khostovan2018}.}. We use the estimator of the ACF proposed by \citet{Landy1993}:
\begin{flalign}
\label{ACF_estimator}
    \omega(\theta)&=\frac{DD(\theta)-2DR(\theta)+RR(\theta)}{RR(\theta)},\\[6pt]
    \varepsilon_{\omega}(\theta)&=\frac{1+\omega(\theta)}{\sqrt{DD_{0}(\theta)}},
\end{flalign}
where $DD(\theta),\ DR(\theta)$, and $RR(\theta)$ are the normalised number counts of galaxy-galaxy, galaxy-random, and random-random pairs whose separations are $\theta$, respectively, and $DD_{0}(\theta)$ is the raw number count of galaxy-galaxy pairs. We use random points with a surface number density of $10\,\mathrm{arcmin}^{-2}$ uniformly distributed over the entire survey footprint and adopt angular bins between $1\arcsec<\theta<3600\arcsec$ with equal intervals on a logarithmic scale. We approximate the ACF with a power-law:
\begin{equation}
\label{omega_model}
    \omega_\mathrm{model}(\theta)=A_{\omega}\theta^{-\beta},
\end{equation}
where $A_{\omega}=\omega(1\arcsec)$ is the amplitude of the ACF. We fix $\beta$ to the fiducial value 0.8 (e.g. \citealp{Peebles1975,Ouchi2003}).

The ACF derived from observational data using equation~\eqref{ACF_estimator} is negatively biased from the true value due to the finite survey area. This bias is known as the integral constraint (IC; \citealp{Groth1977}) and depends on the survey geometry. The IC is estimated using random points \citep{Roche1999}:
\begin{flalign}
    \omega_\mathrm{obs}(\theta)& = \omega_\mathrm{true}(\theta) - \mathrm{IC},\\[6pt]
    \mathrm{IC}& = \frac{\sum_{\theta} RR(\theta)\cdot \omega_\mathrm{model}(\theta)}{\sum_{\theta} RR(\theta)}=\frac{\sum_{\theta} RR(\theta)\cdot A_{\omega}\theta^{-\beta}}{\sum_{\theta} RR(\theta)},
\end{flalign}
where $\omega_\mathrm{obs}$ is the ACF derived from the observational data and $\omega_\mathrm{true}$ is the true ACF. We separately calculate the IC for each of the four DUD fields and find that the ICs are almost the same, $\sim0.0021A_{\omega}$.

We derive the best-fitted amplitude of the ACF over the entire area by minimising $\chi^{2}$:
\begin{equation}
    \chi^{2}=\sum_{j=\mathrm{field}}\sum_{i}\frac{[A_{\omega}\theta_{i}^{-\beta}-({\omega_\mathrm{obs,j}(\theta_{i})+IC_{j}})]^{2}}{\varepsilon_{\omega,j}^{2}(\theta_{i})}
\end{equation}
We use data points at $40\arcsec<\theta<3600\arcsec$, excluding those at $\theta<40\arcsec$ to avoid the contribution of the one-halo term. We also derive the $1\sigma$ error of $A_{\omega}$ from the covariance matrix of the fit.

We then define the spatial two-point correlation function $\xi_\mathrm{g}(r)$:

\begin{equation}
    \xi_\mathrm{g}(r) = \left(\frac{r}{r_{0}}\right)^{-\gamma},
\end{equation}
where $r_{0}$ is the correlation length and $\gamma=1+\beta = 1.8$ is the slope of the power-law. We obtain $\xi_\mathrm{g}(r)$ from $\omega(\theta)$ via the inverse Limber transform \citep{Peebles1980,Efstathiou1991}.

From $\xi_\mathrm{g}(r)$, we calculate the linear bias parameter of galaxies $b_\mathrm{g}$ at a large scale:
\begin{equation}
\label{bias}
    b_\mathrm{g} = \sqrt{\frac{\xi_\mathrm{g}\left(r=8\, \mathrm{cMpc}/ {\it h_\mathrm{100}}\right)}{\xi_\mathrm{DM}\left(r=8\, \mathrm{cMpc}/{\it h_\mathrm{100}}\right)}},
\end{equation}
where $\xi_\mathrm{DM}(r)$ is the spatial correlation function of dark matter, calculated from the matter power spectrum based on the \citet{Eisenstein1998} model. Finally, the $b_\mathrm{g}$ value is converted into the average halo mass using the relation between $b_\mathrm{g}$ and the peak height given in \citet{Tinker2010}. For these calculations, we use a python toolkit for cosmological calculations called \texttt{COLOSSUS} \citep{Diemer2018}.

\subsection{Estimated halo masses}
\label{sec:halo_mass}
\begin{figure*}
	\includegraphics[width=2\columnwidth]{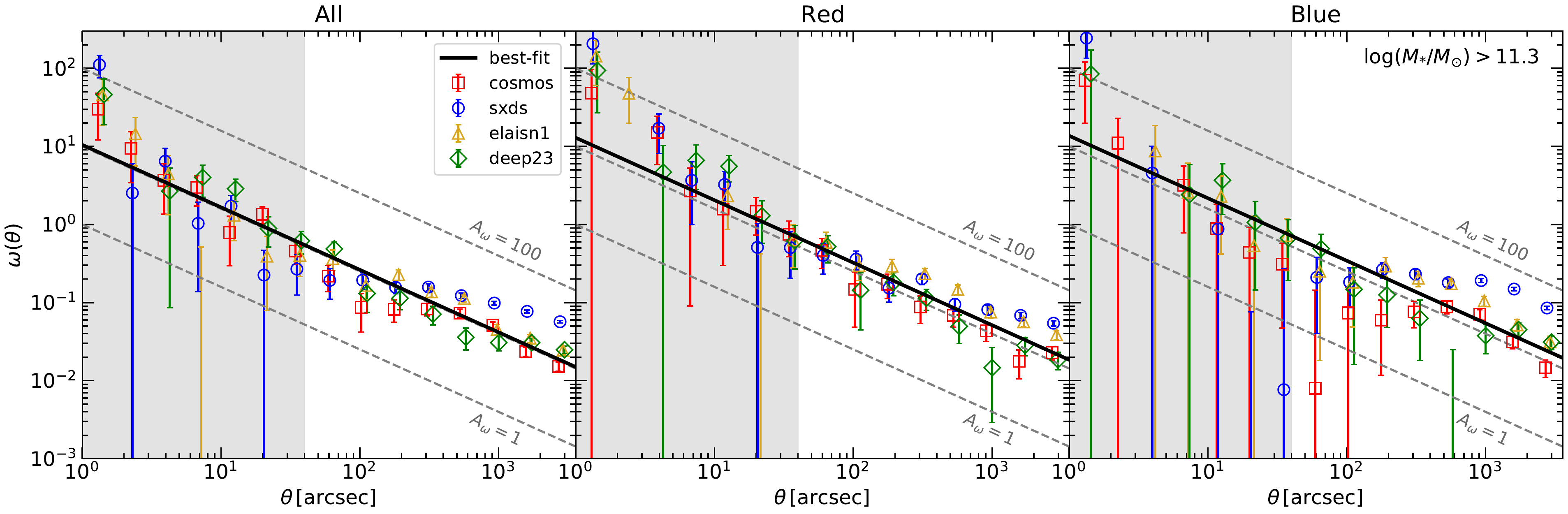}
    \caption{The measured ACFs with the IC correction for all galaxies (left), RGs (middle), and BGs (right) with $\log(M_{*}/M_{\odot})>11.3$. Red squares, blue circles, yellow triangles, and green diamonds are the ACFs in the COSMOS, SXDS, ELAIS-N1, and DEEP2-3 fields, respectively. Each data point is slightly offset along the horizontal axis for clarity. Black solid lines are the best-fit models derived from the simultaneous fit of the observed ACFs. Grey shades show the angular ranges not used for the fitting ($\theta<40\arcsec$). Grey dashed lines are power-law models with $A_\mathrm{\omega}=100$, $10$, and $1$.}
    \label{fig:acf}
\end{figure*}

We select galaxies more massive than a range of stellar mass limit $M_{*,\mathrm{lim}}$, $10.8\leq \log(M_{*,\mathrm{lim}}/M_{\odot})\leq 11.5$, as central galaxy candidates of PC cores. The derived quantities from the clustering analysis are summarised in Table~\ref{tab:clustering}. As an example, we show the measured ACFs with the IC correction in the four DUD fields and the best-fit model for the sample with $\log(M_{*,\mathrm{lim}}/M_{\odot})=11.3$ in Fig.~\ref{fig:acf}. The observed ACFs show a field-to-field variance. In particular, the ACF in the SXDS field clearly exceeds those in the other fields at large scales, suggesting the existence of a tens-megaparsec-scale structure in this field.

\begin{figure}
	\includegraphics[width=\columnwidth]{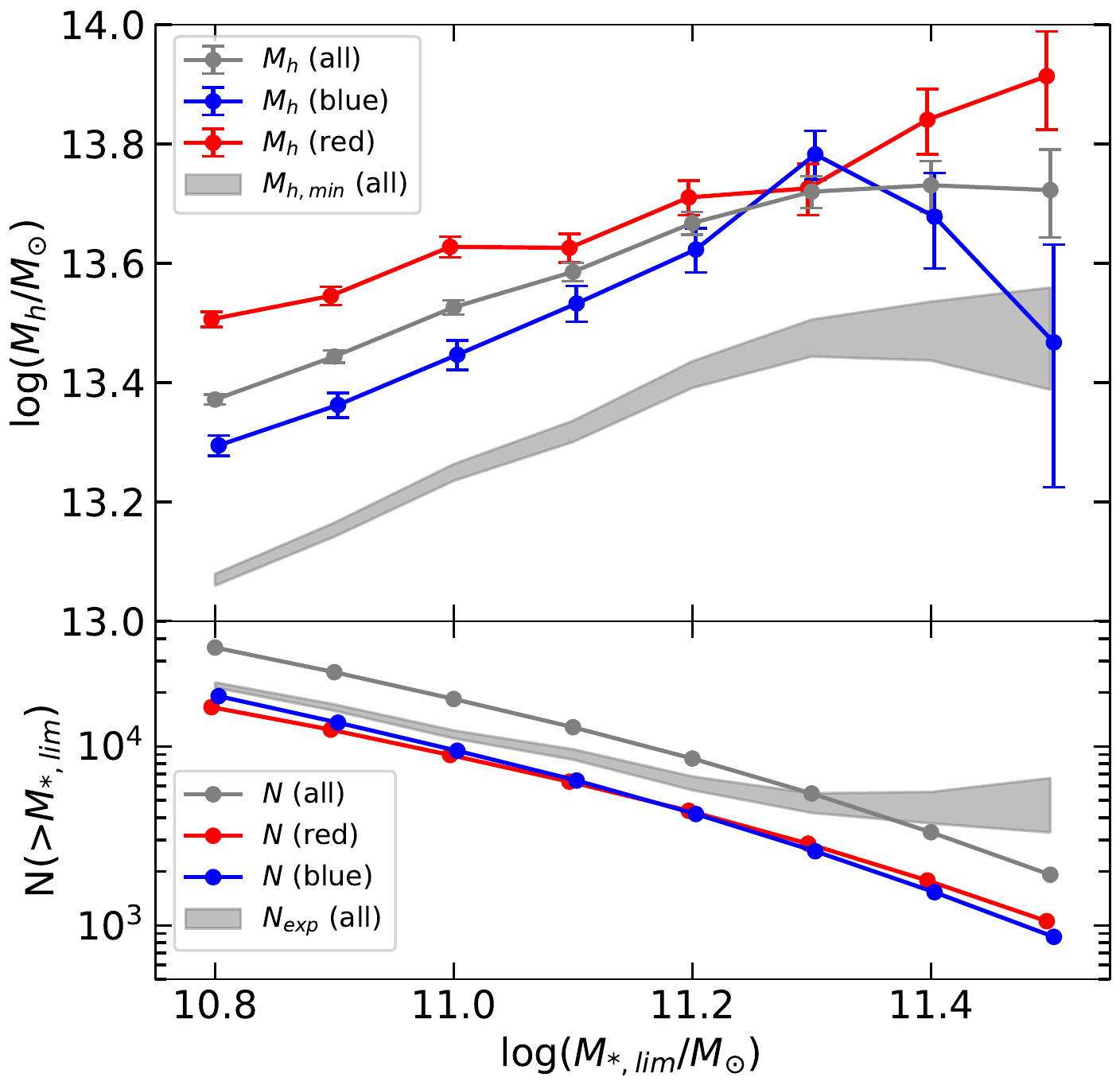}
    \caption{\textit{Top}: Estimated halo masses of all galaxies (grey circles connected by a line), RGs (red), and BGs (blue). A grey shade is the minimum halo mass estimated by equation~\eqref{eq:Mh_min}. \textit{Bottom}: The numbers of galaxies in individual categories more massive than $M_{*,\mathrm{lim}}$, with the same colour coding as the top panel. A grey shade is the expected number of haloes corresponding to the minimum halo mass calculated by the halo mass function.}
    \label{fig:halo_mass}
\end{figure}

We show the estimated average host halo masses in the top panel of Fig.~\ref{fig:halo_mass}. The halo masses of all galaxies, RGs, and BGs are separately plotted as grey, red, and blue solid lines, respectively. In the bottom panel, the numbers of galaxies in individual categories are shown. As a general trend, halo mass increases with the stellar mass limit, with a relatively large uncertainty at $\log(M_{*,\mathrm{lim}}/M_{\odot})\geq11.4$.

\citet{Ishikawa2020} have measured the ACFs of $0.3<z<1.4$ galaxies with different stellar mass limits using \textsc{MIZUKI}'s photo-\textit{z} sample. They have obtained $A_\mathrm{\omega}=\omega(1\arcsec)\sim12$ and $\gamma \sim1.77$ for galaxies with $\log(M_{*}/M_{\odot})\gtrsim11.3$ at $1.1<z<1.4$ (see their Table~3), which agrees with ours. \citet{Cheema2020} have estimated halo masses of massive quiescent galaxies at $z\sim1.6$ from the ACFs. They have found the host halo masses of quiescent galaxies with $\log(M_{*}/M_{\odot})\gtrsim11.3(11.5)$ to be $\log(M_\mathrm{h}/M_{\odot})\sim13.8(14.3)$, roughly consistent with our result for RGs.

In addition to the average halo mass, we also calculate the minimum halo mass in the sample corresponding to each stellar mass limit. An observed bias parameter is a weighted mean of the biases of haloes with different masses, $b(M)$. Therefore, considering the abundance matching between galaxies and haloes and assuming a one-to-one correspondence between them, we calculate the minimum halo mass $M_\mathrm{h, min}$ as follows:

\begin{equation}
\label{eq:Mh_min}
    b_\mathrm{obs} = \frac{ \int^{\infty}_{M_{\mathrm{h,min}}} b(M)\, \frac{dn(M)}{dM}\, dM }{ \int^{\infty}_{M_{\mathrm{h,min}}} \frac{dn(M)}{dM}\, dM },
\end{equation}
where $\frac{dn(M)}{dM}$ is the halo mass function. Here, we adopt the \citet{Sheth1999} halo mass function. Moreover, we calculate the expected number of haloes more massive than $M_\mathrm{h,min}$ which exist in our survey volume, by integrating the halo mass function from $M_\mathrm{h,min}$ to infinity. The derived minimum halo mass and expected number of haloes are shown in the top and bottom panels of Fig.~\ref{fig:halo_mass}, respectively. 

The minimum mass is smaller than the average mass by $\sim0.25\,\mathrm{dex}$ at the entire mass range. At $\log(M_{*,\mathrm{lim}}/M_{\odot})\leq11.2$, there exist more galaxies, and hence haloes, than expected. Since less massive haloes are more abundant, this suggests that the halo mass of these galaxies is overestimated. On the contrary, at $\log(M_{*,\mathrm{lim}}/M_{\odot})\geq11.4$, the number of haloes detected is lower than expected, suggesting underestimation of halo masses. At $\log(M_{*,\mathrm{lim}}/M_{\odot})=11.3$, both the average and minimum halo masses are large enough as PC cores, and the expected number of haloes agrees with that of galaxies. We thus regard galaxies with a stellar mass of $\log(M_{*,\mathrm{lim}}/M_{\odot})\geq11.3$ as candidates of the central galaxies of PC cores.

\begin{table*}
\centering
    \caption{The derived physical quantities from the clustering analysis.}
    \label{tab:clustering}
	\begin{tabular}{cccccccc}
		\hline
		limiting mass & $N$ & $z_\mathrm{ave}$ & $A_{\omega}$ & $r_{0}$ & $b_\mathrm{g}$ & halo mass & minimum halo mass \\
		$\log(M_{*,\mathrm{lim}}/M_{\odot})$ & \# &  &  & cMpc &  & $\log(M_{\mathrm{h}}/M_{\odot})$ & $\log(M_{\mathrm{h,min}}/M_{\odot})$ \\
		\hline
        \multicolumn{8}{c}{all}\\
        \hline
        $10.8$ &$35637$ &$1.22$ &$8.13^{+0.10}_{-0.10}$ & $14.36^{+0.10}_{-0.10}$ & $4.44^{+0.03}_{-0.03}$ & $13.37^{+0.01}_{-0.01}$ & $13.07_{-0.01}^{+0.01}$ \\  [2pt]
        $10.9$ &$25963$ &$1.23$ &$9.09^{+0.14}_{-0.14}$ & $15.30^{+0.13}_{-0.13}$ & $4.71^{+0.04}_{-0.04}$ & $13.44^{+0.01}_{-0.01}$ & $13.15_{-0.01}^{+0.01}$ \\  [2pt]
        $11.0$ &$18392$ &$1.23$ &$10.46^{+0.20}_{-0.20}$ & $16.47^{+0.18}_{-0.18}$ & $5.03^{+0.05}_{-0.05}$ & $13.53^{+0.01}_{-0.01}$ & $13.25_{-0.01}^{+0.01}$ \\  [2pt]
        $11.1$ &$12791$ &$1.23$ &$11.59^{+0.29}_{-0.29}$ & $17.39^{+0.24}_{-0.25}$ & $5.29^{+0.07}_{-0.07}$ & $13.59^{+0.02}_{-0.02}$ & $13.32_{-0.02}^{+0.02}$ \\  [2pt]
        $11.2$ &$8550$ &$1.23$ &$13.43^{+0.44}_{-0.44}$ & $18.80^{+0.34}_{-0.34}$ & $5.68^{+0.09}_{-0.09}$ & $13.67^{+0.02}_{-0.02}$ & $13.41_{-0.02}^{+0.02}$ \\  [2pt]
        $11.3$ &$5452$ &$1.24$ &$14.93^{+0.69}_{-0.69}$ & $19.81^{+0.50}_{-0.51}$ & $5.97^{+0.14}_{-0.14}$ & $13.72^{+0.03}_{-0.03}$ & $13.48_{-0.03}^{+0.03}$ \\  [2pt]
        $11.4$ &$3311$ &$1.24$ &$15.45^{+1.14}_{-1.14}$ & $20.06^{+0.81}_{-0.84}$ & $6.05^{+0.22}_{-0.23}$ & $13.73^{+0.04}_{-0.04}$ & $13.49_{-0.05}^{+0.05}$ \\  [2pt]
        $11.5$ &$1917$ &$1.25$ &$15.55^{+1.98}_{-1.98}$ & $19.98^{+1.38}_{-1.46}$ & $6.05^{+0.37}_{-0.40}$ & $13.72^{+0.07}_{-0.08}$ & $13.48_{-0.09}^{+0.08}$ \\  [2pt]
        \hline
        \multicolumn{8}{c}{red}\\
        \hline
        $10.8$ &$16560$ &$1.18$ &$11.39^{+0.23}_{-0.23}$ & $15.90^{+0.18}_{-0.18}$ & $4.79^{+0.05}_{-0.05}$ & $13.51^{+0.01}_{-0.01}$ & - \\  [2pt]
        $10.9$ &$12393$ &$1.18$ &$12.43^{+0.31}_{-0.31}$ & $16.45^{+0.22}_{-0.23}$ & $4.94^{+0.06}_{-0.06}$ & $13.55^{+0.02}_{-0.02}$ & - \\  [2pt]
        $11.0$ &$8945$ &$1.18$ &$14.71^{+0.43}_{-0.43}$ & $17.71^{+0.28}_{-0.29}$ & $5.27^{+0.08}_{-0.08}$ & $13.63^{+0.02}_{-0.02}$ & - \\  [2pt]
        $11.1$ &$6347$ &$1.18$ &$15.06^{+0.60}_{-0.60}$ & $17.67^{+0.39}_{-0.40}$ & $5.25^{+0.10}_{-0.11}$ & $13.63^{+0.02}_{-0.02}$ & - \\  [2pt]
        $11.2$ &$4363$ &$1.18$ &$17.70^{+0.88}_{-0.88}$ & $19.12^{+0.52}_{-0.54}$ & $5.64^{+0.14}_{-0.14}$ & $13.71^{+0.03}_{-0.03}$ & - \\  [2pt]
        $11.3$ &$2858$ &$1.18$ &$18.43^{+1.36}_{-1.36}$ & $19.42^{+0.78}_{-0.81}$ & $5.72^{+0.21}_{-0.21}$ & $13.73^{+0.04}_{-0.04}$ & - \\  [2pt]
        $11.4$ &$1780$ &$1.18$ &$22.65^{+2.22}_{-2.22}$ & $21.78^{+1.16}_{-1.21}$ & $6.36^{+0.30}_{-0.32}$ & $13.84^{+0.05}_{-0.06}$ & - \\  [2pt]
        $11.5$ &$1055$ &$1.20$ &$25.29^{+3.81}_{-3.81}$ & $23.58^{+1.91}_{-2.05}$ & $6.86^{+0.50}_{-0.54}$ & $13.91^{+0.07}_{-0.09}$ & - \\  [2pt]
        \hline
        \multicolumn{8}{c}{blue}\\
        \hline
        $10.8$ &$19077$ &$1.26$ &$7.57^{+0.19}_{-0.19}$ & $13.64^{+0.19}_{-0.19}$ & $4.30^{+0.05}_{-0.05}$ & $13.29^{+0.02}_{-0.02}$ & - \\  [2pt]
        $10.9$ &$13570$ &$1.27$ &$8.47^{+0.27}_{-0.27}$ & $14.48^{+0.25}_{-0.26}$ & $4.55^{+0.07}_{-0.07}$ & $13.36^{+0.02}_{-0.02}$ & - \\  [2pt]
        $11.0$ &$9447$ &$1.27$ &$9.86^{+0.39}_{-0.39}$ & $15.62^{+0.34}_{-0.34}$ & $4.89^{+0.10}_{-0.10}$ & $13.45^{+0.02}_{-0.03}$ & - \\  [2pt]
        $11.1$ &$6444$ &$1.28$ &$11.60^{+0.57}_{-0.57}$ & $16.92^{+0.46}_{-0.47}$ & $5.27^{+0.13}_{-0.13}$ & $13.53^{+0.03}_{-0.03}$ & - \\  [2pt]
        $11.2$ &$4187$ &$1.29$ &$14.07^{+0.88}_{-0.88}$ & $18.47^{+0.63}_{-0.65}$ & $5.73^{+0.18}_{-0.18}$ & $13.62^{+0.04}_{-0.04}$ & - \\  [2pt]
        $11.3$ &$2594$ &$1.30$ &$19.52^{+1.42}_{-1.42}$ & $21.65^{+0.86}_{-0.89}$ & $6.64^{+0.24}_{-0.25}$ & $13.78^{+0.04}_{-0.04}$ & - \\  [2pt]
        $11.4$ &$1531$ &$1.32$ &$17.19^{+2.36}_{-2.36}$ & $19.64^{+1.45}_{-1.55}$ & $6.10^{+0.41}_{-0.43}$ & $13.68^{+0.07}_{-0.09}$ & - \\  [2pt]
        $11.5$ &$862$ &$1.32$ &$13.02^{+4.05}_{-4.05}$ & $16.22^{+2.64}_{-3.04}$ & $5.16^{+0.75}_{-0.88}$ & $13.47^{+0.16}_{-0.24}$ & - \\  [2pt]
        \hline
	\end{tabular}
\end{table*}

\subsection{Halo mass evolution}
\label{sec:mass_evol}
We infer the mass growth of PC core candidates using the IllustrisTNG \citep{Nelson2019}.

The IllustrisTNG project is a series of cosmological magnetohydrodynamical simulations of galaxy formation including various baryon physics \citep{Pillepich2018a, Weinberger2017}. We use results from TNG300-1, which has the high mass resolution ($\sim4\times10^{7}\,M_{\odot}/h$ for dark matter particles) and the large volume (side length of $205\, \mathrm{cMpc}/h$), and hence is suitable for investigating rare objects like PCs. See also IllustisTNG presentation papers for a detailed description \citep{Naiman2018,Springel2018,Pillepich2018c,Marinacci2018,Nelson2018}.

We use the halo (group) and galaxy (subhalo) catalogues as well as merger trees. We extract two subsamples from the halo catalogue mimicking observationally selected PC core candidates. One is a halo-mass matched sample consisting of haloes with a mass of $\log(M_\mathrm{h}/M_{\odot})=13.72\pm 0.1$, and the other is a stellar mass matched sample consisting of central galaxies with $\log(M_{*}/M_{\odot})\geq11.3$. We identify 119 and 887 haloes in the halo-mass matched sample and the stellar-mass matched sample, respectively. We then track the mass growth of haloes in each subsample down to $z=0$ by tracing the merger trees.

\begin{figure*}
	\includegraphics[width=2\columnwidth]{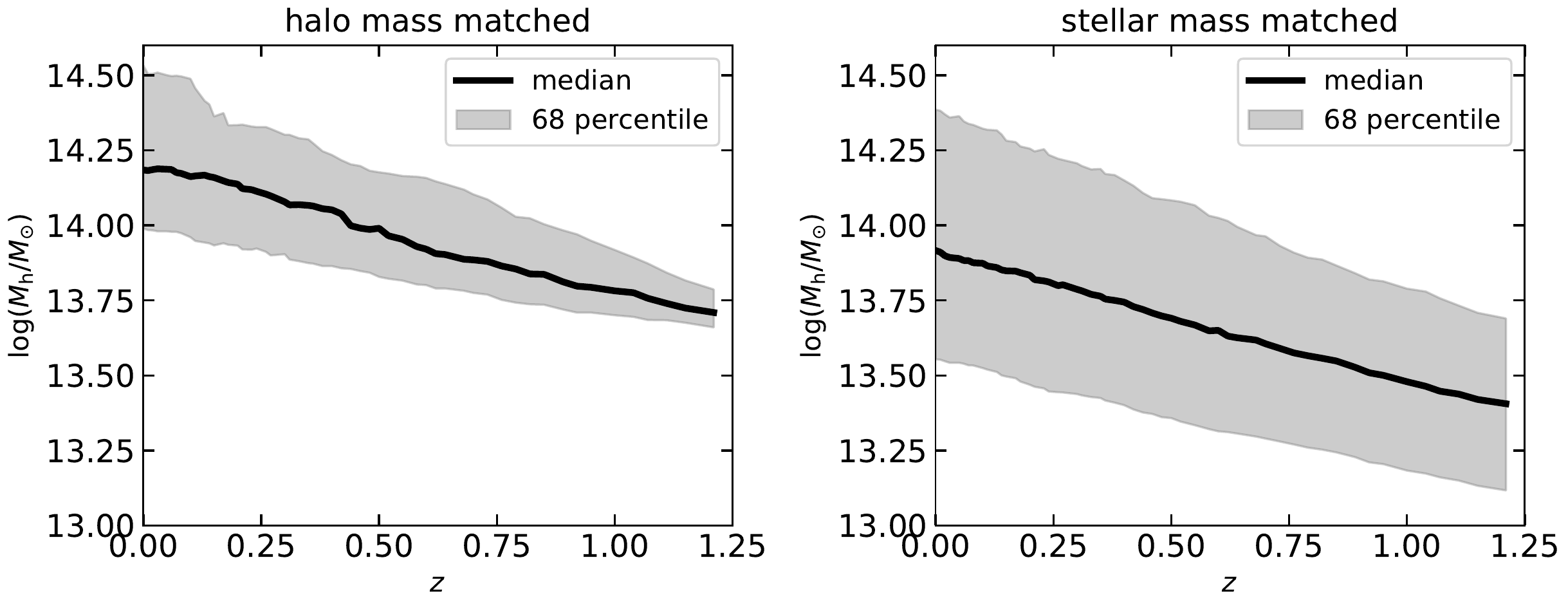}
    \caption{The redshift evolution of halo mass for the halo-mass matched sample (left) and the stellar-mass matched sample (right). The median values and 68 percentiles are shown by solid lines and shades, respectively.}
    \label{fig:halo_evolution}
\end{figure*}

The redshift evolution of halo mass for the two subsamples is shown in Fig.~\ref{fig:halo_evolution}. The medians and 68 percentiles of halo mass are plotted as solid lines and shades. The median halo masses of the two subsamples at $z=0$ reach the cluster mass regime ($\log(M_{*}/M_{\odot})\gtrsim14$), although the stellar-mass matched sample shows a slightly smaller mass and a larger scatter. We also test whether the result of the halo-mass matched sample changes if we use the minimum halo mass $M_{\mathrm{h,min}}$ instead of the average value to select haloes: $\log(M_\mathrm{h}/M_{\odot})\geq \log(M_\mathrm{h,min}/M_{\odot})=13.47$. We find only slight changes: the median halo mass becomes smaller by $\sim 0.05\,\mathrm{dex}$, and the scatter becomes larger by $\sim 0.17\,\mathrm{dex}$. In any case, galaxies more massive than $10^{11.3}\,M_{\odot}$ are hosted by very massive haloes which are likely to grow into clusters by $z=0$. Considering these facts, we regard such massive galaxies as the central galaxies of PC cores. In what follows, PC core central galaxies denote galaxies more massive than $10^{11.3}\,M_{\odot}$.

We emphasise that the descendant masses of the halo mass matched sample are consistent with Fornax-like or Virgo-like clusters ($14<\log(M_\mathrm{h}/M_{\odot})<15$ at $z=0$; \citealp{Chiang2013}) rather than Coma-like clusters ($\log(M_\mathrm{h}/M_{\odot})>15$ at $z=0$). This suggests that our PC cores are likely to be progenitors of typical clusters in the present-day Universe.

\subsection{Overdensity around PC core candidates}
\label{sec:overdensity}
In the previous subsection, we confirm our PC core candidates in terms of halo mass. Here, we present another test of the candidates. PCs are essentially overdense regions that extend to at least several $\mathrm{pMpc}$ beyond the virial radius of the core regions \citep{Chiang2013, Muldrew2015}. Therefore, our PC core candidates are expected to be surrounded by such large-scale overdensities. We examine the overdensity profiles around PC core central galaxies as follows.

First, we extract all galaxies with $10<\log(M_{*}/M_{\odot})<11$ in cylindrical regions around the central galaxies of the cores with a photo-\textit{z} difference of $\Delta z=\pm0.15$. To sample field galaxies, we create five times as many random points as the central galaxies with the same redshift distribution as the central galaxies. We also select galaxies around random points in the same manner as above. Then, we count the number of selected galaxies as a function of the projected distance from centrals with a correction of masked areas. Finally, we derive overdensity $\delta$ as:

\begin{equation}
\label{eq:overdensity}
    \delta(<r) = \frac{N_{\mathrm{PC}}(<r)}{N_{\mathrm{field}}(<r)}-1,
\end{equation}
where $N_{\mathrm{PC}(<r)}$ and $N_{\mathrm{field}(<r)}$ are the galaxy number counts within radius $r$ around central galaxies and random points, respectively. Here, we calculate overdensities for red and blue centrals separately.

\begin{figure}
	\includegraphics[width=\columnwidth]{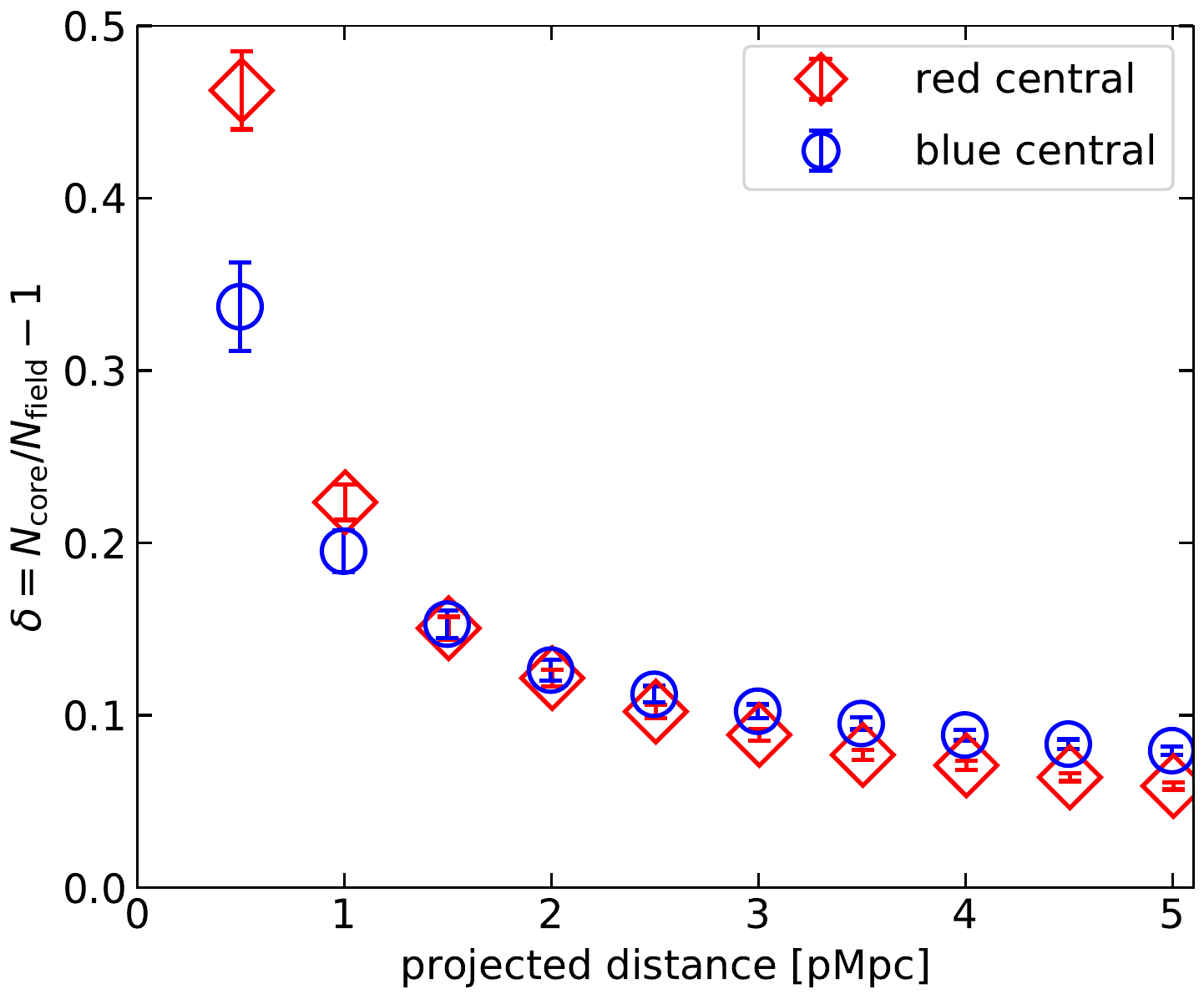}
    \caption{The overdensity profiles around PC core candidates with red central galaxies (red diamonds) and blue ones (blue circles) as a function of projected distance. The error bars are the Poisson errors of number counts. There are clear overdensities both small ($<1\,\mathrm{pMpc}$) and large ($>1\,\mathrm{pMpc}$) scales.}
    \label{fig:overdensity}
\end{figure}

The overdensity profiles averaged over the four DUD fields are shown in Fig.~\ref{fig:overdensity}. We find a clear overdensity around both red and blue central galaxies. The observed overdensities are very significant, especially at a small scale ($<1\,\mathrm{pMpc}$), where core regions would exist. Interestingly, galaxies are more concentrated around red centrals than blue centrals, although their average host halo masses are comparable. The overdensities extend to a larger scale ($>1\,\mathrm{pMpc}$) beyond the expected virial radius of the cores. These overdensity profiles are consistent with a picture that highly overdense regions, PC cores, are surrounded by milder large-scale overdensities, i.e., the rest of the PC regions. We note that an overdensity is also seen up to several $\mathrm{pMpc}$ in the differential radial profile for both red and blue centrals.

With these overdensity profiles and halo mass estimates, we conclude that massive galaxies with $\log(M_{*}/M_{\odot})>11.3$ are likely to trace PC cores. These PC cores are more common systems than very massive ($\log(M_\mathrm{h}/M_{\odot})>14$) haloes at $z>1$ (e.g., \citealp{vanderBurg2020,Cheema2020}).

\section{Properties of member galaxies of PC cores}
\label{sec:properties}
To examine the star formation activity in the PC cores, we calculate the SMF and the red galaxy fraction. These two quantities are complementary since the SMF reflects the cumulative (past) star formation history while the red fraction shows the differential (current) star formation activity. Here, we assume that PC cores are spheres with a radius of $0.5\,\mathrm{pMpc}$ centred on central galaxies. This radius is close to $r_{200}$ of haloes with $\log(M_\mathrm{h}/M_{\odot})\sim13.7$ expected by a spherical collapse model and where the projected average overdensity is very significant. Since photo-\textit{z}'s have relatively large uncertainties, we need to subtract the contribution of field galaxies to calculate the SMF and the red fraction in PC cores.

\subsection{Field subtraction and the field stellar mass function}
\label{sec:field_subtraction}
For each PC core, we count all galaxies except for the central galaxy in the cylindrical region with a radius of $\Delta r=0.5\,\mathrm{pMpc}$ and a line-of-sight length $\Delta z=\pm0.15$, avoiding counting the same galaxies multiple times. Since field galaxies contaminate these cylindrical regions, we perform field subtraction in a similar manner to \citet{Ando2020} as follows.

First, we calculate the SMFs of field galaxies by dividing the galaxy sample of $\log(M_{*}/M_{\sun})\geq 9.0$ over $0.85<z<1.65$ into redshift bins, ${z_{i}}$, with a width of $\Delta z=0.05$, referred to as $\Phi_{\mathrm{field}}(z_{i})$. For each redshift bin, we also compute the total cosmic volume occupied by the cylindrical regions around the PC core centrals, ${V_\mathrm{c}(z_{i})}$, taking into account the overlap of the cylinders. Then, we estimate the total number of contamination galaxies falling within the cylindrical regions as a function of stellar mass, to be $\sum_{i}{V_\mathrm{c}(z_{i})\cdot \Phi_{\mathrm{field}}(z_{i})}$. Finally, we subtract these expected number counts of contaminants from the raw counts around the PC core centrals.

To directly compare the SMFs of PC cores with those of field galaxies, we need to calculate the field SMFs reflecting different redshift distributions of all, red, and blue centrals. We take weighted means of $\Phi_{\mathrm{field}}(z_{i})$ for a given central galaxy category, $j=\{\mathrm{all,red,blue}\}$, as:

\begin{equation}
\label{phi_field}
    \Phi_\mathrm{field}^{j} = \frac{\sum_{i}{V^{j}_\mathrm{c}(z_{i})\cdot \Phi_{\mathrm{field}}(z_{i})}}{\sum_{i}{V^{j}_\mathrm{c}(z_{i})}},
\end{equation}
where $V^{j}_\mathrm{c}(z_{i})$ is the total volume at ${z_{i}}$ occupied by cylinders around central galaxies of the given category.

\subsection{The stellar mass function}
\label{sec:smf}
\begin{figure*}
	\includegraphics[width=2\columnwidth]{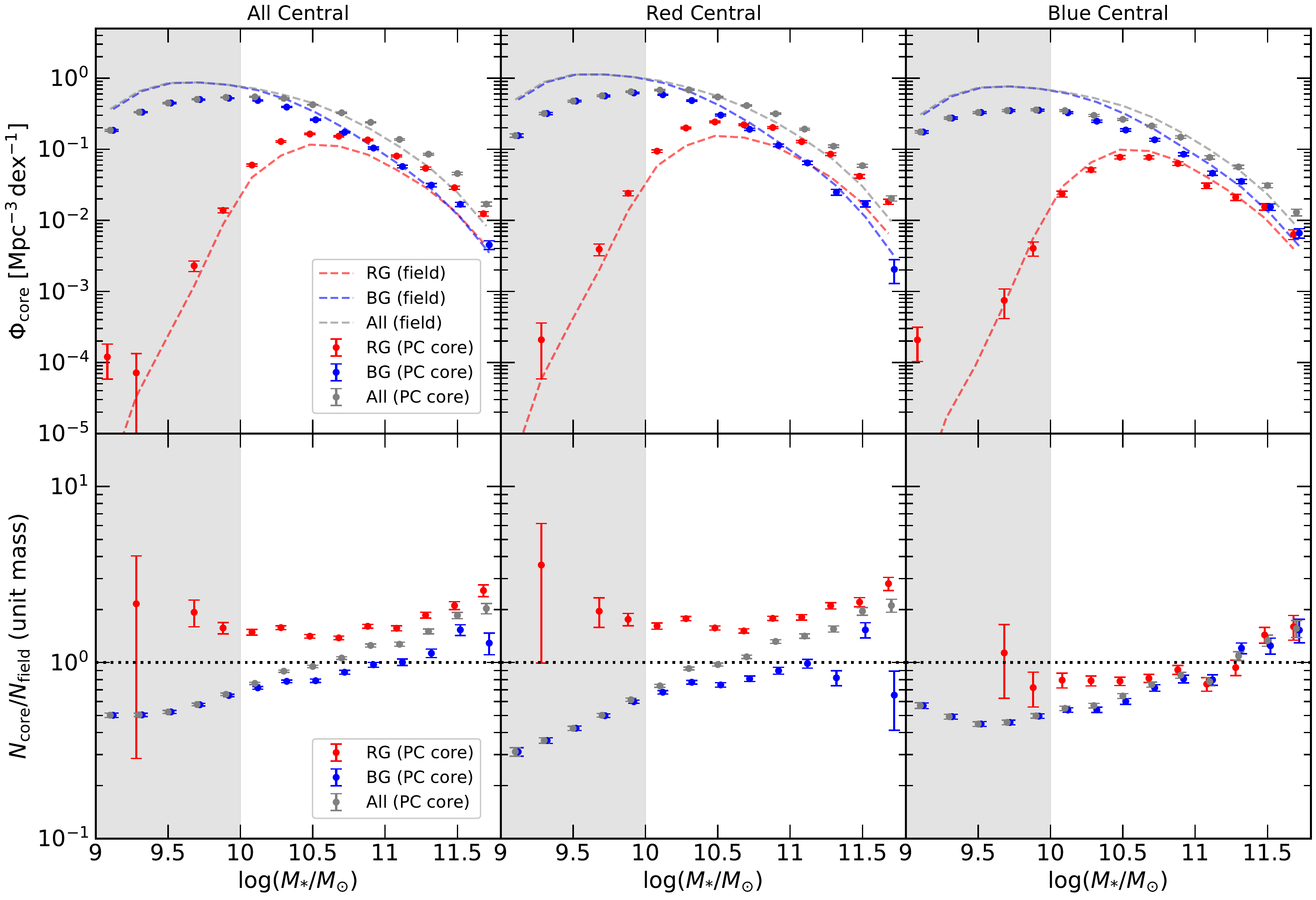}
    \caption{\textit{Top panel}: The stellar mass functions (SMFs) of galaxies in the PC cores with all (left), red (middle) and blue (right) centrals. Dashed lines show the SMFs of field galaxies. Grey, blue, and red colours mean the SMFs of all galaxies, RGs and BGs, respectively. Detection incompleteness has not been corrected. Grey shades are unreliable mass ranges due to incompleteness.
    \textit{Bottom panel}: Same as top panels but divided by the field SMFs and normalised by total mass using equation~\eqref{eq:smf_unitmass}. Black dotted lines indicate unity.}
    \label{fig:SMF}
\end{figure*}

The SMFs of galaxies in the PC cores are shown in the top panels of Fig.~\ref{fig:SMF}, where the left, middle, and right panels are, respectively, for all cores, cores with red centrals, and cores with blue centrals. Grey, red, and blue dots refer to the SMFs of all galaxies, RGs, and BGs, respectively. We do not correct them for completeness, although the results are almost unchanged even if the detection completeness correction is applied.

To discuss the shapes of the SMFs in the PC cores, we take the ratio of the SMF of PC core galaxies to that of field galaxies for each star formation category as \citet{Ando2020} have done. We normalise the ratio by total mass as:
\begin{equation}
\label{eq:smf_unitmass}
    \frac{N_\mathrm{core}}{N_\mathrm{field}} = \frac{\Phi_\mathrm{core}}{\Phi_\mathrm{field}}\cdot \mathrm{norm}=\frac{\Phi_\mathrm{core}}{\Phi_\mathrm{field}} \cdot \frac{\rho_\mathrm{crit}\Omega_\mathrm{m}V_\mathrm{core}}{M_\mathrm{core}},
\end{equation}
where $\rho_\mathrm{crit}$ is the critical density of the universe, $V_\mathrm{core}$ is the average comoving volume of cores, and $M_\mathrm{core}$ is the average halo mass of cores. Using this ratio, we can make a non-parametric comparison without the correction of selection completeness and detection completeness, and free from uncertainties in parametric modelling like the Schechter function. 

The results are plotted in the bottom panels of Fig.~\ref{fig:SMF}. For visual comparison, we also show the re-scaled field SMFs, ${\Phi_\mathrm{field}}/\mathrm{norm}$, in the top panels of Fig.~\ref{fig:SMF}. We find that the ratios for all galaxies and BGs increase with stellar mass, i.e., the SMFs in PC cores are more top-heavy than in the field. On the other hand, the shapes of SMFs for RGs are similar regardless of environment. These trends are common among the three central galaxy categories. We also find that the ratio of the SMFs for BGs around red centrals has a steeper slope than around blue centrals, especially at the low-mass range ($\log(M_{*}/M_{\odot})\lesssim10$), suggesting a more top-heavy SMF in PC cores with red centrals.

The top-heavy SMFs in the PC cores for all galaxies and BGs are consistent with simulations by \citet{Lovell2018,Muldrew2018}. They have found that the formation of high-mass galaxies in PC cores is enhanced than in the field by earlier formation of high-mass haloes and higher merger rates, while low-mass galaxies are reduced in number by mergers and/or tidal disruption in core regions. Some other observations also support top-heavy SMFs at $z\gtrsim1$. By a similar analysis to this study, \citet{Ando2020} have found that PC core candidates at $z\sim2$ have a top-heavy SMF. From deep photometric and spectroscopic observations of cluster galaxies at $1<z<1.4$, \citet{vanderBurg2020} have shown that the SMF of all galaxies in clusters is more top-heavy than that of the field. They have also pointed out identical SMF shapes for star-forming and quiescent populations between the two environments. This trend for the star-forming population is different from our result, i.e. the SMF for BGs is also more top-heavy in PC cores.

It has been reported that PCs at $z>2$ show a sign of concentration of high-mass galaxies and sometimes lack low-mass galaxies \citep{Cooke2014,Shimakawa2018,Ito2020}. Because PC cores grow through accretion from the rest of the PC region, the top-heavy SMFs in cores observed at $1<z<1.5$ may partly originate from the stellar mass distribution of accreted galaxies skewed toward higher masses. A fraction of these massive satellites will merge with their central galaxies and promote the formation of brightest cluster galaxies \citep{Sawicki2020}, i.e. outstandingly giant elliptical galaxies found in matured clusters.

We note that the ratios of the SMFs are below unity over most of the mass range except for RGs around all and red centrals. This might suggest that the galaxy formation per unit mass is less efficient in PC core regions. However, this interpretation is inconsistent with the previous studies that argue more efficient galaxy formation in high-redshift clusters or PC cores at least at $\log(M_{*}/M_{\odot})>10$ \citep{vanderBurg2013,vanderBurg2018,Ando2020}. Another possible cause is a systematic error in the normalisation defined in equation~\eqref{eq:smf_unitmass} due to underestimation of $V_\mathrm{core}$ and/or overestimation of $M_\mathrm{core}$. In any case, the relative galaxy formation efficiency is higher for higher-mass galaxies in the PC core environment.

\subsection{The red galaxy fraction}
\label{sec:fr}
\begin{figure*}
	\includegraphics[width=2\columnwidth]{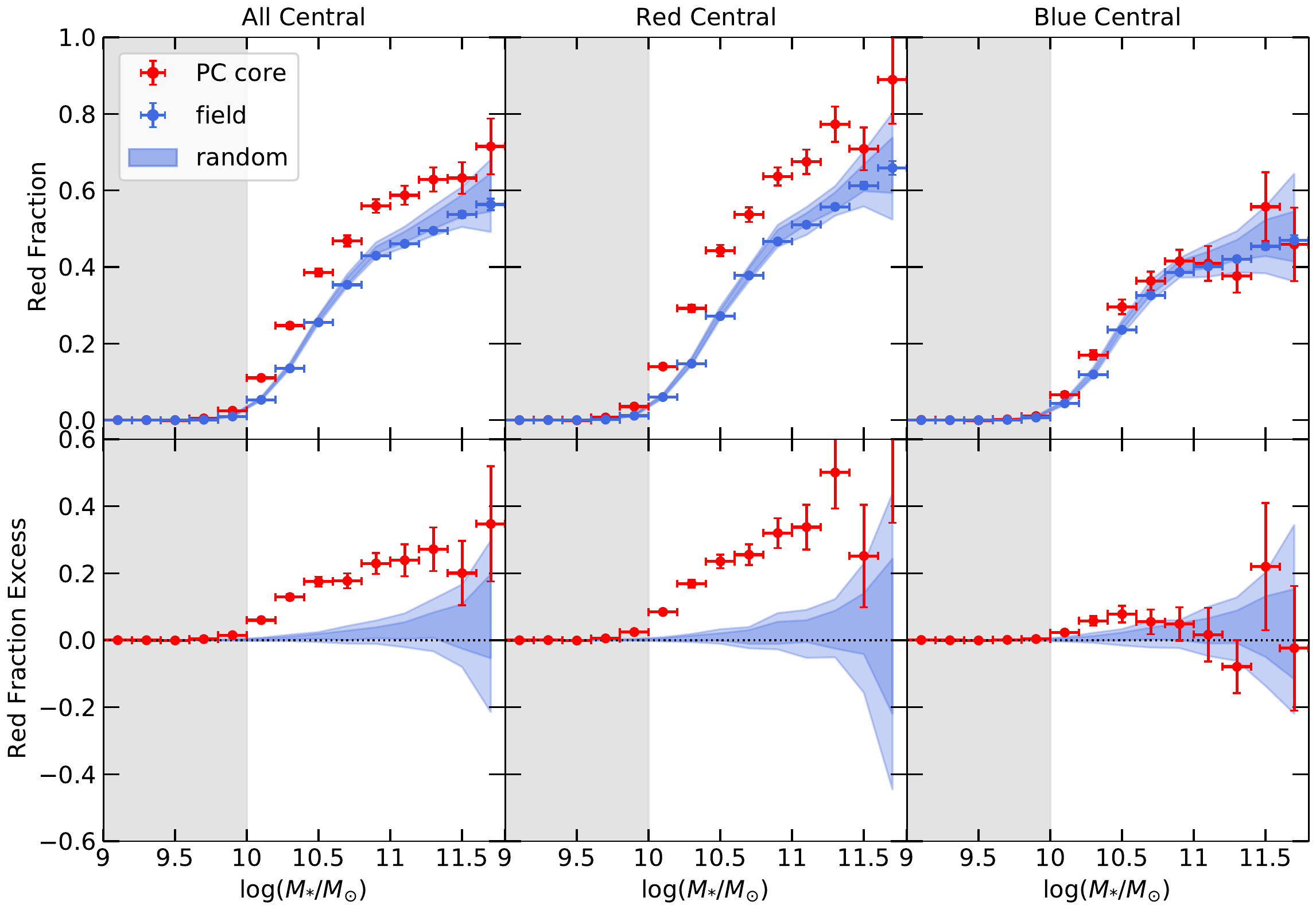}
    \caption{\textit{Top}: The red galaxy fraction ($f_\mathrm{r}$) in PC cores with all (left), red (middle), and blue centrals (right). The $f_\mathrm{r}$ in PC cores and that in the field are plotted as red and blue symbols, respectively. Dark (light) blue shades show the 68th (95th) percentile of the $f_\mathrm{r}$ distribution measured around random points to evaluate the significance of the excess of the observed $f_\mathrm{r}$ in PC cores. Grey shades are unreliable mass ranges due to incompleteness.
    \textit{Bottom}: The red fraction excess, $\mathrm{RFE}=(f_\mathrm{r}^\mathrm{core}-f_\mathrm{r}^\mathrm{field})/(1-f_\mathrm{r}^\mathrm{field})$, in PC cores. The meaning of the symbols is the same as the top panel.}
    \label{fig:fr}
\end{figure*}

We measure the red fraction in the PC cores to examine the environmental dependence of star formation activity. The red fraction, $f_\mathrm{r}$, is defined as:
\begin{equation}
    f_\mathrm{r}=\frac{N_\mathrm{red}}{N_\mathrm{all}},
\end{equation}
where $N_\mathrm{all}$ and $N_\mathrm{red}$ are the numbers of all galaxies and RGs, respectively. For PC cores, we use the number counts after field subtraction. The red fractions averaged over the four DUD fields are shown in the top panels of Fig.~\ref{fig:fr}. The red fractions in the PC cores and the corresponding field are shown as red and blue points, respectively. In both PC and field regions, the red fraction increases with stellar mass. The red fractions for all centrals and red centrals are systematically higher than those of the field, while only a small excess is seen for blue centrals.

These results appear to suggest a concentration of red galaxies in the PC cores. However, similar results would be obtained if field galaxies along the sightlines of PC cores happen to have a higher red fraction than the cosmic average due to cosmic variance because we only consider the averaged field population for field subtraction. We check whether such a cosmic variance of field galaxies can reproduce the measured red fractions. We create random points with the same number and redshift distribution as the central galaxies and measure the red fraction using cylinders around them. In each DUD field, we iterate this procedure one hundred times. Then, we derive one hundred sets of red fractions averaged over the four DUD fields. We show the 68th (95th) percentiles of the red fraction derived in this way as dark (light) blue shades in Fig.~\ref{fig:fr}. The red fraction around red centrals exceeds the 95th percentile over most of the mass range, while that around blue centrals nearly overlaps with it. This suggests that the excess at least around red centrals is real. The red fractions averaged over all stellar masses are summarised in Table~\ref{tab:fr}.

To quantify the excess red fraction in PC cores, we calculate the red fraction excess (RFE) defined as:
\begin{equation}
    \label{eq:rfe}
    \mathrm{RFE}=\frac{f_\mathrm{r}^\mathrm{core}-f_\mathrm{r}^\mathrm{field}}{1-f_\mathrm{r}^\mathrm{field}} = 1-\frac{f_\mathrm{b}^\mathrm{core}}{f_\mathrm{b}^\mathrm{field}},
\end{equation}
where $f_\mathrm{r}^\mathrm{core}$ and $f_\mathrm{r}^\mathrm{field}$ ($f_\mathrm{b}^\mathrm{core}$ and $f_\mathrm{b}^\mathrm{field}$) are the red (blue) fraction of galaxies in PC cores and in the field. Since the red fraction indicates the dominance of quiescent galaxies in a given environment, the RFE indicates how much environmental effects additionally quench star-forming galaxies.

We show the RFE in the PC cores in the bottom panel of Fig.~\ref{fig:fr}. The RFE for all centrals is positive and increases with stellar mass at $\log(M_{*}/M_{\odot})>10$. These trends are significant beyond what can be explained by the statistical errors or the cosmic variance of field galaxies. This means that environmental quenching depends on stellar mass and is more effective for higher-mass galaxies. At $z\lesssim1$, environmental quenching is independent of stellar mass (e.g., \citealp{Peng2010}), while stellar-mass dependent environmental quenching has been found at $z\gtrsim1$ \citep{Balogh2014,vanderBurg2020,Reeves2021}. Our result shows that this trend continues at $z>1$ at a statistically high significance.

A similar but somewhat stronger trend is seen around red centrals. Very interestingly, the RFE around blue centrals is much smaller than those of all and red centrals and almost within the uncertainty due to the cosmic variance of field galaxies. This is a sign of galactic conformity in PC cores at $z>1$. We return to this point in \S~\ref{sec:conformity}.

We compare our RFE with previous measurements of the quiescent fraction excess (QFE)\footnote{The definition of the QFE is obtained from equation~\eqref{eq:rfe} by replacing the red galaxy fraction$f_\mathrm{r}$ with the quiescent galaxy fraction $f_\mathrm{q}$. There are various terminologies in the literature to denote the same quantity as the QFE: transition fraction \citep{vandenBosch2008}, environmental quenching efficiency (e.g., \citealp{Peng2010}), and conversion fraction (e.g., \citealp{Balogh2016}).} for similar halo mass groups.\footnote{We note that the RFE and the QFE are similar but, strictly speaking, different quantities. However, these two quantities can reasonably be compared since our selection of RG agrees with the widely used QG selection criteria (see \S~\ref{sec:class}).} \citet{Sarron2021} have searched for galaxy groups at $z<2.5$ as stellar-mass overdensities using a \textit{K}-selected sample of the REFINE survey and found about 120 reliable groups at $1<z<1.5$ with $\log(M_\mathrm{h}/M_{\odot})\sim13.7$. They have obtained $0.1\lesssim \mathrm{QFE}\lesssim0.2$ for galaxies with $10.25<\log(M_{*}/M_{\odot})<11$. \citet{Reeves2021} have found $\mathrm{QFE}\sim 0.27$ for galaxies with $\log(M_{*}/M_{\odot})>10$ in groups at $1<z<1.5$ with $13.65<\log(M_\mathrm{h}/M_{\odot})<13.9$ conformed by the X-ray and spectroscopic observations of the GOGREEN survey.
The RFE of our study is comparable to that of \citet{Sarron2021} but slightly smaller to that of \citet{Reeves2021}: $0.172\pm0.005$ around all centrals for galaxies with $\log(M_{*}/M_{\odot})>10$ (see Table~\ref{tab:fr}). There are several issues to consider. \citet{Sarron2021} have focused on the central regions ($<0.5\times R_{200}$) of their groups when computing the QFE. Because the QFE decreases with group-centric-radius \citep{vanderBurg2018,vanderBurg2020}, the measured RFE may become smaller if outer regions of haloes are included as in our case. In fact, when we calculate the RFE within a smaller radius of $0.3\,\mathrm{pMpc}$, the RFE increases slightly to be $0.216\pm0.007$ around all centrals for galaxies with $\log(M_{*}/M_{\odot})>10$, still consistent with the QFE of \citet{Sarron2021} (see also Table~\ref{tab:fr}). The QFE also has halo mass dependence, with lower QFEs for lower mass haloes \citep{Nantais2016,Reeves2021}. As shown in Fig.~\ref{fig:halo_mass}, our PC core sample contains less massive haloes at least down to $\log(M_\mathrm{h,min}/M_{\odot})\sim13.5$. This could lead to the lower RFE compared to the QFE of \citet{Reeves2021} who mainly use on average a more massive sample (see their Table~1).

\begin{table*}
\centering
    \caption{The red galaxy fraction and the red fraction excess.}
    \label{tab:fr}
	\begin{tabular}{cccccc}
		\hline
		$\log(M_{*}/M_{\odot})$ & $f_\mathrm{r}^\mathrm{core}$ ($<0.5\,\mathrm{pMpc}$) & $f_\mathrm{r}^\mathrm{core}$ ($<0.3\,\mathrm{pMpc}$) & $f_\mathrm{r}^\mathrm{field}$ & RFE ($<0.5\,\mathrm{pMpc}$) & RFE ($<0.3\,\mathrm{pMpc}$) \\
		\hline
        \multicolumn{6}{c}{all}\\
        \hline
        $[10.0,10.5)$ & $0.203 \pm 0.004$ & $0.233 \pm 0.005$ & $0.111 \pm 0.001$ & $0.103 \pm 0.005$ & $0.136 \pm 0.006$ \\  [2pt]
        $[10.5,\inf)$ & $0.526 \pm 0.008$ & $0.551 \pm 0.010$ & $0.384 \pm 0.001$ & $0.231 \pm 0.014$ & $0.271 \pm 0.016$ \\  [2pt]
        $[10.0,\inf)$ & $0.348 \pm 0.004$ & $0.382 \pm 0.005$ & $0.213 \pm 0.001$ & $0.172 \pm 0.005$ & $0.216 \pm 0.007$ \\
        \hline
        \multicolumn{6}{c}{red}\\
        \hline
        $[10.0,10.5)$ & $0.244 \pm 0.006$ & $0.279 \pm 0.007$ & $0.122 \pm 0.001$ & $0.139 \pm 0.007$ & $0.179 \pm 0.008$ \\  [2pt]
        $[10.5,\inf)$ & $0.605 \pm 0.011$ & $0.647 \pm 0.013$ & $0.417 \pm 0.001$ & $0.323 \pm 0.019$ & $0.396 \pm 0.022$ \\  [2pt]
        $[10.0,\inf)$ & $0.408 \pm 0.006$ & $0.455 \pm 0.007$ & $0.230 \pm 0.001$ & $0.232 \pm 0.008$ & $0.292 \pm 0.009$ \\
        \hline
        \multicolumn{6}{c}{blue}\\
        \hline
        $[10.0,10.5)$ & $0.141 \pm 0.007$ & $0.147 \pm 0.008$ & $0.099 \pm 0.001$ & $0.047 \pm 0.008$ & $0.053 \pm 0.009$ \\  [2pt]
        $[10.5,\inf)$ & $0.390 \pm 0.014$ & $0.346 \pm 0.015$ & $0.346 \pm 0.001$ & $0.068 \pm 0.022$ & $-0.004 \pm 0.023$ \\  [2pt]
        $[10.0,\inf)$ & $0.253 \pm 0.007$ & $0.238 \pm 0.008$ & $0.192 \pm 0.001$ & $0.076 \pm 0.009$ & $0.057 \pm 0.010$ \\
        \hline
	\end{tabular}
\end{table*}

\section{Discussion}
\subsection{Comparison with literature RFEs}
\begin{figure}
	\includegraphics[width=\columnwidth]{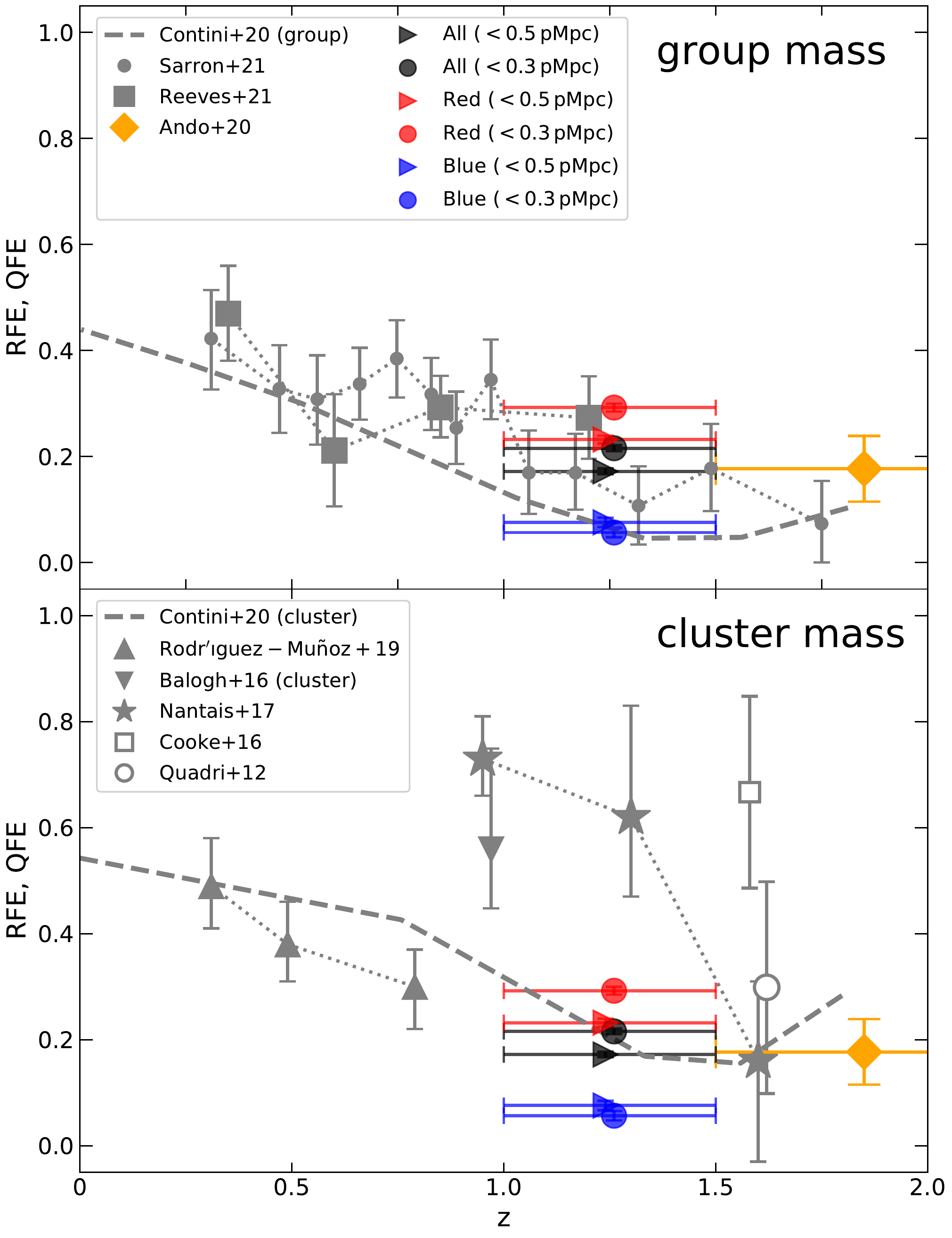}
    \caption{\textit{Top}: The RFEs of this study and the QFEs of groups ($\log(M_\mathrm{h}/M_{\odot})<14$) in the literature as a function of redshift. Black, red, and blue rightward triangles (circles) are the RFEs around all, red, and blue centrals measured within $0.5\, (0.3)\,\mathrm{pMpc}$. The orange diamond is the QFE measured in the PC cores at $1.5<z<3$ presented in \citet{Ando2020}. The other grey symbols show the QFEs in the literature. Dots and squares are the QFEs presented in \citet{Sarron2021} and \citet{Reeves2021}, respectively. A grey dashed line shows the QFE calculated for groups in an analytic galaxy formation model in \citep{Contini2020}. 
    \textit{Bottom}: The same as the top panel, but for the RFEs and the QFEs of clusters ($\log(M_\mathrm{h}/M_{\odot})>14$). Upward triangles, a downward triangle, stars, an open square, and an open circle show the QFEs presented in \citet{Rodriguez2019}, \citet{Balogh2016}, \citet{Nantais2017}, \citet{Cooke2016}, and \citet{Quadri2012}, respectively. Open symbols indicate the QFEs for individual clusters. The grey dashed line shows the QFEs for clusters predicted by \citet{Contini2020}. We also plot the results of this study and \citet{Ando2020} again for comparison.
    }
    \label{fig:rfe_qfe}
\end{figure}

In \S~\ref{sec:fr}, we compare the RFEs in PC cores with those of groups with similar halo masses and redshifts. In this subsection, we further investigate the redshift and halo mass dependence of the RFE. Fig.~\ref{fig:rfe_qfe} shows our RFE estimates ($\log(M_{*}/M_{\odot})>10$) and literature QFE estimates. In the top panel, we plot the QFEs in group mass environment, i.e.  $\log(M_\mathrm{h}/M_{\odot})\lesssim14$. \citet{Ando2020} have searched for PC cores in a similar manner to this study and calculated the QFE at $\log(M_{*}/M_{\odot})>10.3$. \citet{Sarron2021} have calculated QFEs at $10.25<\log(M_{*}/M_{\odot})<11$ in groups identified as stellar mass overdensities. \citet{Reeves2021} have analysed their own group sample as well as that of previous studies \citep{Giodini2012,Omand2014} to obtain QFEs at $\log(M_{*}/M_{\odot})>10$. \citet{Contini2020} have used an analytic galaxy formation model to calculate QFEs for group mass ($13.5<\log(M_\mathrm{h}/M_{\odot})<14$) and cluster mass ($\log(M_\mathrm{h}/M_{\odot})>14.2$) haloes. We have to note that the group finding method, the definition of red or quiescent galaxies, and the stellar mass range are different among these studies.

The RFEs and QFEs are found to mildly increase with decreasing redshift. We find the RFEs in PC cores at $1<z<1.5$ to be comparable to the QFE of PC cores at $1.5<z<3$ \citep{Ando2020}. This suggests violent quenching does not occur in typical PC cores before $z\sim1$, because the PC cores in this study are on the expected mass evolution track of those in \citet{Ando2020}. This might be a reflection of the fact that the whole structure of PCs has not significantly collapsed before $z\sim1$ \citep{Muldrew2015,Nantais2017}.

We also compile QFEs in cluster mass environment, i.e. $\log(M_\mathrm{h}/M_{\odot})\gtrsim 14$, as shown in the bottom panel of Fig.~\ref{fig:rfe_qfe}. \citet{Rodriguez2019} have calculated QFEs at $\log(M_{*}/M_{\odot})>10$ for 24 X-ray detected clusters at $0.2<z<0.9$. For \citet{Balogh2016}, we plot QFE at $\log(M_{*}/M_{\odot})=10.5$ for nine spectroscopically confirmed clusters at $0.85<z<1.25$. \citet{Nantais2017} have shown QFEs at $\log(M_{*}/M_{\odot})>10.3$ in 14 spectroscopically confirmed clusters at $0.9\lesssim z\lesssim 1.6$. As for individual cluster cases at $z\sim1.6$, we show the QFEs of \citet{Quadri2012} and \citet{Cooke2016} calculated at $\log(M_{*}/M_{\odot})>10$ and $10<\log(M_{*}/M_{\odot})<10.7$, respectively. Our RFE estimates are also plotted for comparison.

\citet{Reeves2021} have argued that halo mass is a primary driver of quenching at a fixed redshift. With a somewhat large scatter, the bottom panel of Fig.\ref{fig:rfe_qfe} shows that some massive clusters have significantly larger QFEs than groups (i.e., PC cores). Such massive haloes, however, are progenitors of the most massive clusters like the Coma cluster ($\log(M_\mathrm{h}/M_{\odot})>15$ at $z=0$; \citealp{Chiang2013}). To discuss what a typical cluster, like the Virgo or Fornax cluster ($14<\log(M_\mathrm{h}/M_{\odot})<15$ at $z=0$), looks like during its formation epoch, we need to observe less massive group-like objects at $z>1$.

\subsection{Galactic conformity}
\label{sec:conformity}
\citet{Weinmann2006} have first reported galactic conformity as a phenomenon that the red (quiescent) galaxy fraction is higher in galaxy clusters with red (quiescent) centrals than those with blue (star-forming) centrals. After the first detection, galactic conformity is widely interpreted as similarities in other properties like specific-SFR \citep{Kauffmann2013}, morphology \citep{Otter2020}, and \ion{H}{I} gas content \citep{Li2021}. Interestingly, conformity signals are also detected up to several to ten $\mathrm{Mpc}$ beyond a single halo scale (e.g., \citealp{Kauffmann2013}). The galactic conformities inside and outside of the halo scale ($\sim 1\,\mathrm{Mpc}$) are termed one-halo and two-halo conformity, respectively. In the following part of this section, we mainly focus on the one-halo conformity in galaxy star formation.

In the low-redshift universe, a conformity signal is detected even if halo masses are fixed (e.g., \citealp{Weinmann2006,Knobel2015}), suggesting that halo mass is not the single parameter to control the building up of the quenched population in a dense environment. Recently, galactic conformity has also been found even at high-redshift beyond $z=1$ out to $z\sim2$ \citep{Hartley2015,Kawinwanichakij2017,Hatfield2017,Alam2020}. As an example, \citet{Hartley2015} have used photometric data in the UKIDSS UDS field to investigate galactic conformity at $0.4<z<1.9$. They have compared the satellite quiescent fraction between star-forming centrals with $\log(M_{*}/M_{\odot})>11$ and quiescent centrals with $10.5<\log(M_{*}/M_{\odot})<11$ to approximately match the host halo masses \citep{Hartley2013}. They have found that the quenched fraction is higher than the field value only around quiescent centrals, suggesting that the halo mass difference is not the exclusive origin of galactic conformity. However, this finding is only marginal or indirect evidence that conformity signals are independent of halo mass since estimating halo masses at high-redshift is much more difficult than at low-redshift.

Many physical origins of the conformity at high-redshift are proposed, such as assembly bias (e.g., \citealp{Hearin2016,Berti2017}), inhibition of gas cooling around massive galaxies \citep{Hartley2015,Kawinwanichakij2016}, and combinations of several quenching processes. In any case, the physical causes of conformity are still unclear at any redshift.

We detect a significant conformity signal around galaxies with $\log(M_{*}/M_{\odot})>11.3$ (see Fig.~\ref{fig:fr}). Importantly, these central galaxies have almost the same halo masses as shown in Fig.~\ref{fig:halo_mass} and Table~\ref{tab:clustering}. This is direct evidence that there exists galactic conformity not just caused by halo mass differences even at $z>1$ at least in high-mass haloes. We note that there is still a possibility that the observed conformity signal is caused by the central galaxies at the highest-stellar-mass range ($\log(M_{*}/M_{\odot})>11.4$) in our sample, where red centrals have higher halo masses than blue centrals. However, the halo-mass errors in this mass range are very large. A larger sample is needed to test this possibility.

To quantify the strength of conformity, we calculate the star formation suppression factor, $\xi_\mathrm{conf}$, introduced in \citet{Knobel2015}:
\begin{equation}
    \xi_\mathrm{conf}=\frac{\mathrm{RFE_{blue}}}{\mathrm{RFE_{red}}},
\end{equation}
where $\mathrm{RFE_{red}}$ and $\mathrm{RFE_{blue}}$ are the RFEs around red and blue centrals, respectively. $\xi_\mathrm{conf}$ indicates how the environment with star-forming centrals suppresses the quenching effect compared to that with quiescent centrals. Usually, $\xi_\mathrm{conf}$ takes zero to unity with smaller values meaning stronger conformity.

\begin{figure}
	\includegraphics[width=\columnwidth]{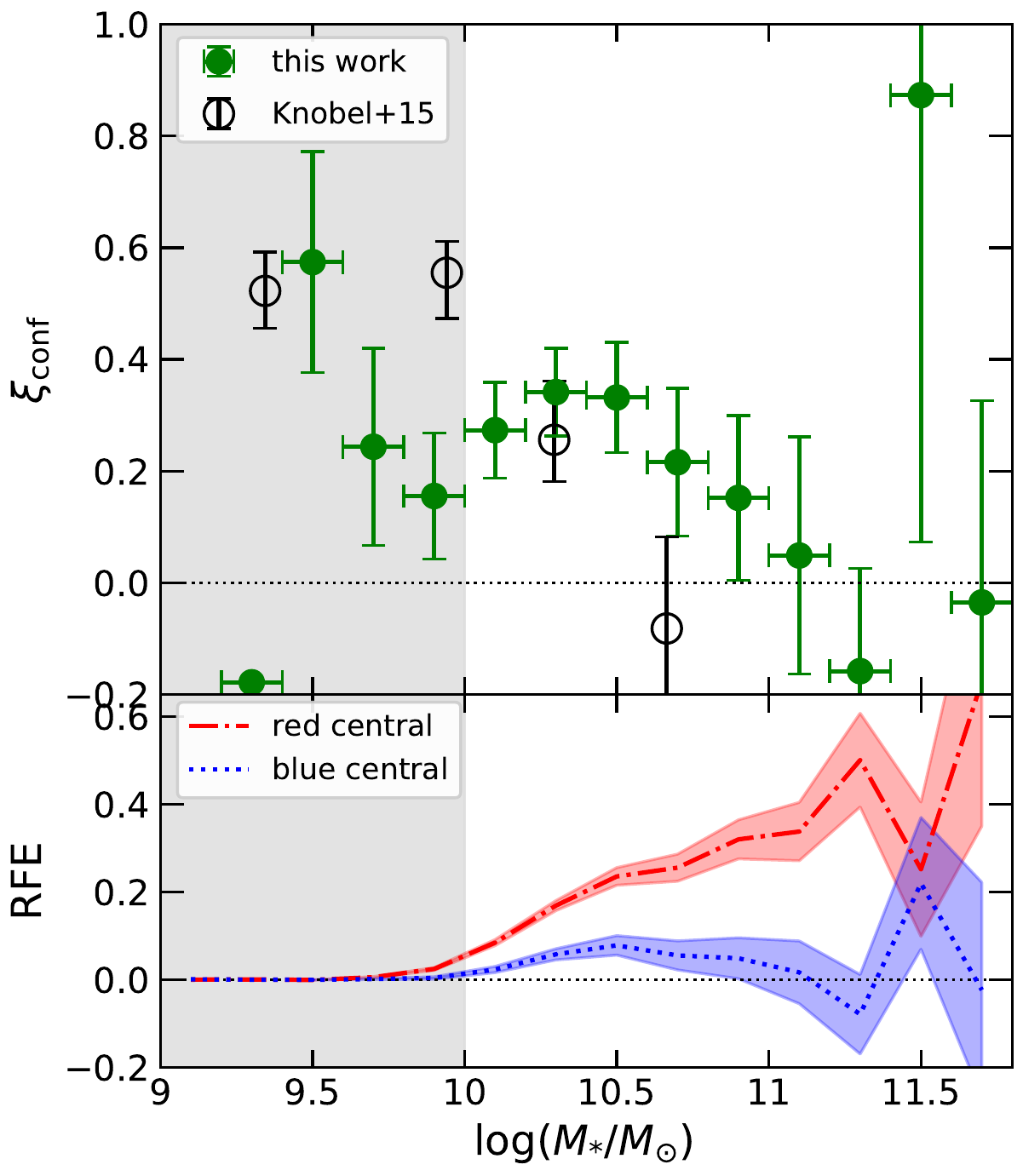}
    \caption{\textit{Top}: The suppression factor of conformity, $\xi_\mathrm{conf}=\mathrm{RFE_\mathrm{blue}}/\mathrm{RFE_\mathrm{red}}$. Green filled circles and black open circles show $\xi_\mathrm{conf}$ of this work and \citet{Knobel2015}, respectively. Grey shades show the unreliable mass ranges. Smaller $\xi_\mathrm{conf}$ means stronger conformity. \textit{Bottom}: The RFEs and their $1\sigma$ ranges in PC cores with red and blue centrals.}
    \label{fig:xi_conf}
\end{figure}

We show $\xi_\mathrm{conf}$ measured within $0.5\,\mathrm{pMpc}$ from centrals together with the RFEs around red and blue centrals in Fig.~\ref{fig:xi_conf}. $\xi_\mathrm{conf}$ takes about 0.05 to 0.35 at $10<\log(M_{*}/M_{\odot})<11.3$, and the value averaged over this mass range is $0.33\pm0.04$. This means that the RFE around red centrals is larger by three to even twenty times than that around blue centrals in each mass bin. \citet{Knobel2015} have used a spectroscopically conformed group sample at $0.01<z<0.06$ and reported an averaged $\xi_\mathrm{conf}$ of $\sim 0.39$, comparable to ours. They have also reported halo mass dependence of $\xi_\mathrm{conf}$ in their Fig.~10. Interestingly, they have shown that conformity is weak ($\xi_\mathrm{conf}\gtrsim 0.7$) for haloes with $\log(M_{\mathrm{h}}/M_{\odot})\sim 13.5$, an opposite result to ours. This may suggest that the conformity strength at a fixed halo mass depends on redshift.

Due to limited data, we cannot make a conclusive discussion about physical processes that cause galactic conformity. However, we can point out a few things that might support one of the physical origins of conformity, namely, assembly bias. Galaxy evolution, including star formation quenching, is accelerated in denser environments, and a higher red (quiescent) fraction can be observed in such regions. We attempt to explain our conformity signal in this context.

As we show in Fig.~\ref{fig:overdensity}, the overdensity profiles around red and blue centrals are different at the halo scale ($<1\,\mathrm{pMpc}$), with a higher overdensity around red centrals. This implies an earlier assembly of galaxies in PC cores with red centrals, resulting in earlier quenching, and thus a higher RFE at a given cosmic time. Moreover, we also find that the ratio of SMFs between PC cores and field regions is different between red and blue centrals as shown in the bottom panels of Fig.~\ref{fig:SMF}. The SMF for all galaxies around red centrals is more top-heavy (i.e., the ratio has a steeper slope) than that around blue centrals. This can be explained, at least partly, by more frequent mergers that result in reducing low-mass galaxies and enhancing the formation of high-mass galaxies. This possibility is supported by the fact that red centrals are located in a locally denser environment than blue centrals. If mergers occur frequently, merger-driven quenching or harassment is enhanced (e.g., \citealp{Moore1998,Peng2010}). In these ways, a different mass assembly history may contribute to building different populations of satellite galaxies between red and blue centrals.

However, this scenario seems to qualitatively conflict with the fact that the halo masses are almost the same between the red and blue centrals because earlier mass assembly around red centrals may result in higher masses of their host haloes. This inconsistency may suggest the presence of other causes that link the star formation activity of centrals and satellites. In any case, to examine whether different assembly histories are the main cause of the conformity seen in the PC cores, it is important to derive the mass weighted-age of satellite galaxies by spectroscopic follow-up observation to determine the formation redshift, as \citet{Webb2020} have done.

\section{Summary}
We have searched for PC cores at $1<z<1.5$ using very wide ($\sim22.2\,\mathrm{deg}$) and deep ($i\sim26.8\,\mathrm{mag}$) optical data from the HSC-SSP which are complete above $\log(M_{*}/M_{\odot})\sim10$. We have defined RGs and BGs using the rest-frame $\mathrm{NUV}-g$ colour and examined the SMF and the red fraction to investigate the quiescence of galaxies in PC cores. The main results are as follows.
\begin{enumerate}
  \item We estimate the halo masses of galaxies more massive than a range of stellar mass limit, $M_\mathrm{*,lim}$, by the two point auto-correlation function measured at $40\arcsec-3600\arcsec$ scale. The average and minimum halo masses of galaxies with stellar masses of $\log(M_{*}/M_{\odot})>11.3$ are $\log(M_\mathrm{h}/M_{\odot})=13.72\pm0.03$ and $\log(M_\mathrm{h,min}/M_{\odot})=13.48\pm0.03$, respectively. The observed number of these galaxies, and thus haloes, matches the expected abundance of haloes more massive than $M_\mathrm{h,min}$ calculated from the halo mass function, implying that the estimated halo mass is correct. We regard these galaxies as the central galaxies of PC cores.
  
  \item To examine the mass growth of these haloes, we extract two halo samples from the IllustrisTNG simulation catalogue mimicking the observationally selected PC core candidates: halo-mass matched sample and stellar-mass matched sample. We then track the mass growth of haloes in each subsample down to $z=0$ by tracing the merger trees and find that they actually grow into the cluster mass regime, $\log(M_\mathrm{h}/M_{\odot})\gtrsim14$, by $z=0$.

  \item We calculate the radial profile of the overdensity around the candidates of PC cores, finding a clear overdensity around both red and blue central galaxies at small ($<1\,\mathrm{pMpc}$) scales as well as at large ($>1\,\mathrm{pMpc}$) scales. Interestingly, at small scales, galaxies are more concentrated around red centrals than blue centrals. These overdensity profiles are consistent with a picture that highly overdense regions, PC cores, are surrounded by milder large-scale overdensities. These overdensity profiles combined with the estimated halo mass suggest that the massive galaxies with $\log(M_{*}/M_{\odot})>11.3$ are PC cores.
  
  \item We calculate the SMFs in the PC cores with all, red, and blue central galaxies separately. The SMFs for all and blue satellites show a top-heavy shape compared to the field SMF, suggesting an enhancement of high-mass galaxy formation due to mergers or the early formation of massive haloes in PC regions, and/or the reduction of low-mass galaxies by mergers and/or tidal disruption. On the other hand, the SMF for red satellites has a comparable shape to that of the field.
  
  \item The red fractions in the PC cores with all and red centrals exceed the field value at $\log(M_{*}/M_{\odot})>10$, and these excesses increase with stellar mass. On the contrary, the red fraction around blue centrals is consistent with the field value within the uncertainty due to the cosmic variance of the field red fraction. The RFE around red centrals is positive while that around blue centrals is almost consistent with zero, suggesting the existence of galactic conformity in PC cores at $z>1$. Since the halo masses of the red and blue centrals are similar, the conformity signal implies that the halo mass is not the single parameter to control galaxy quenching in PC cores. Although the physical origin of the conformity in our sample is not obvious, a more top-heavy SMF and a more significant overdensity around red centrals than blue ones might suggest that a different assembly history at least partly causes the conformity. To test this hypothesis, we need to estimate the stellar mass weighted ages of satellite galaxies by spectroscopy.
  
  \item The QFE mildly increases with decreasing redshift if halo masses are limited to the group mass regime ($\log(M_{*}/M_{\odot})<14$). Galaxies in PC cores may not be violently quenched at least before $z\sim1$, implying that quenching will be more accelerated after this epoch when the whole structures of PCs start to collapse into cores. At a fixed redshift ($1\lesssim z\lesssim1.5$), some cluster-mass haloes show much larger QFEs than group-mass ones. Although the scatter is not small, this suggests that halo mass is a considerable drivers of quenching.
  Note that high-redshift clusters are likely progenitors of the most massive and rare clusters and not those of typical ones. To reveal the formation history of typical clusters, one needs to investigate high-redshift groups.

\end{enumerate}

As a future prospective, we will extend the redshift range of our PC core search up to $z\sim3$ using multi wavelength photometry from \textrm{NUV} to near-infrared provided by the HSC-SSP collaboration. With near-infrared data, we can classify QGs free from dusty SFGs and hence do the similar analysis to this study with higher reliability.

Not only photometric data, but also spectroscopic data are important to confirm PC core candidates and determine physical parameters such as stellar age. The PC core sample constructed in this study provides good targets for upcoming large spectroscopic observation campaigns such as the Subaru Prime Focus Spectrograph (PFS) survey \citep{Takada2014}. Combined with these data, our approach will help to understand cluster galaxy formation and evolution across a wide range of cosmic time and in a wide parameter space.

\section*{Acknowledgements}
We appreciate the anonymous referee for constructive comments. We also thank Shogo Ishikawa and Naoaki Yamamoto for valuable comments and discussions. MA acknowledges support from Iwadare Scholarship Foundation and Japan Science and Technology Agency (JST) Support for Pioneering Research Initiated by the Next Generation (SPRING), Grant Number JPMJSP2108. KS acknowledges support from Japan Society for the Promotion of Science (JSPS) KAKENHI Grant Number JP19K03924. RM acknowledges support from JSPS KAKENHI Grant Number 21H04490 and the Institute for AI and Beyond of the University of Tokyo. KI acknowledges support from JSPS KAKENHI Grant Number 20J12461.

The Hyper Suprime-Cam (HSC) collaboration includes the astronomical communities of Japan and Taiwan, and Princeton University. The HSC instrumentation and software were developed by the National Astronomical Observatory of Japan (NAOJ), the Kavli Institute for the Physics and Mathematics of the Universe (Kavli IPMU), the University of Tokyo, the High Energy Accelerator Research Organisation (KEK), the Academia Sinica Institute for Astronomy and Astrophysics in Taiwan (ASIAA), and Princeton University. Funding was contributed by the FIRST program from the Japanese Cabinet Office, the Ministry of Education, Culture, Sports, Science and Technology (MEXT), JSPS, JST, the Toray Science  Foundation, NAOJ, Kavli IPMU, KEK, ASIAA, and Princeton University.

This paper is based on data collected at the Subaru Telescope and retrieved from the HSC data archive system, which is operated by Subaru Telescope and Astronomy Data Centre (ADC) at NAOJ. Data analysis was in part carried out with the cooperation of Centre for Computational Astrophysics (CfCA) at NAOJ.  We are honoured and grateful for the opportunity of observing the Universe from Maunakea, which has the cultural, historical and natural significance in Hawaii.

This paper makes use of software developed for Vera C. Rubin Observatory. We thank the Rubin Observatory for making their code available as free software at \url{http://pipelines.lsst.io/}. 

The Pan-STARRS1 Surveys (PS1) and the PS1 public science archive have been made possible through contributions by the Institute for Astronomy, the University of Hawaii, the Pan-STARRS Project Office, the Max Planck Society and its participating institutes, the Max Planck Institute for Astronomy, Heidelberg, and the Max Planck Institute for Extraterrestrial Physics, Garching, The Johns Hopkins University, Durham University, the University of Edinburgh, the Queen’s University Belfast, the Harvard-Smithsonian Centre for Astrophysics, the Las Cumbres Observatory Global Telescope Network Incorporated, the National Central University of Taiwan, the Space Telescope Science Institute, the National Aeronautics and Space Administration under grant No. NNX08AR22G issued through the Planetary Science Division of the NASA Science Mission Directorate, the National Science Foundation grant No. AST-1238877, the University of Maryland, Eotvos Lorand University (ELTE), the Los Alamos National Laboratory, and the Gordon and Betty Moore Foundation.

Part of this study is based on data products from observations made with ESO Telescopes at the La Silla Paranal Observatory under ESO programme ID 179.A-2005 and on data products produced by TERAPIX and the Cambridge Astronomy Survey Unit on behalf of the UltraVISTA consortium.

We use the following open source software packages for our analysis: \texttt{numpy} \citep{numpy:2011}, \texttt{pandas} \citep{pandas:2010}, \texttt{scipy} \citep{scipy:2001}, \texttt{astropy} \citep{astropy:2013,astropy:2018} and \texttt{matplotlib} \citep{matplotlib:2007}.

\section*{Data Availability}
The HSC-SSP PDR3 is available in the data release site at \url{https://hsc-release.mtk.nao.ac.jp/doc/}. The COSMOS2015 catalogue is available in the COSMOS survey page at \url{ftp://ftp.iap.fr/pub/from_users/hjmcc/COSMOS2015/}. The datasets of the IllustrisTNG are also available in the IllustrisTNG project page at \url{https://www.tng-project.org/data/}. See also the data release paper \citep{Nelson2019}.




\bibliographystyle{mnras}
\bibliography{pCLcore_hsc_1} 



\appendix
\section{Uncertainties in photometric redshift estimates over the full redshift range}
\label{app:photoz_prec}
\begin{figure*}
	\includegraphics[width=2\columnwidth]{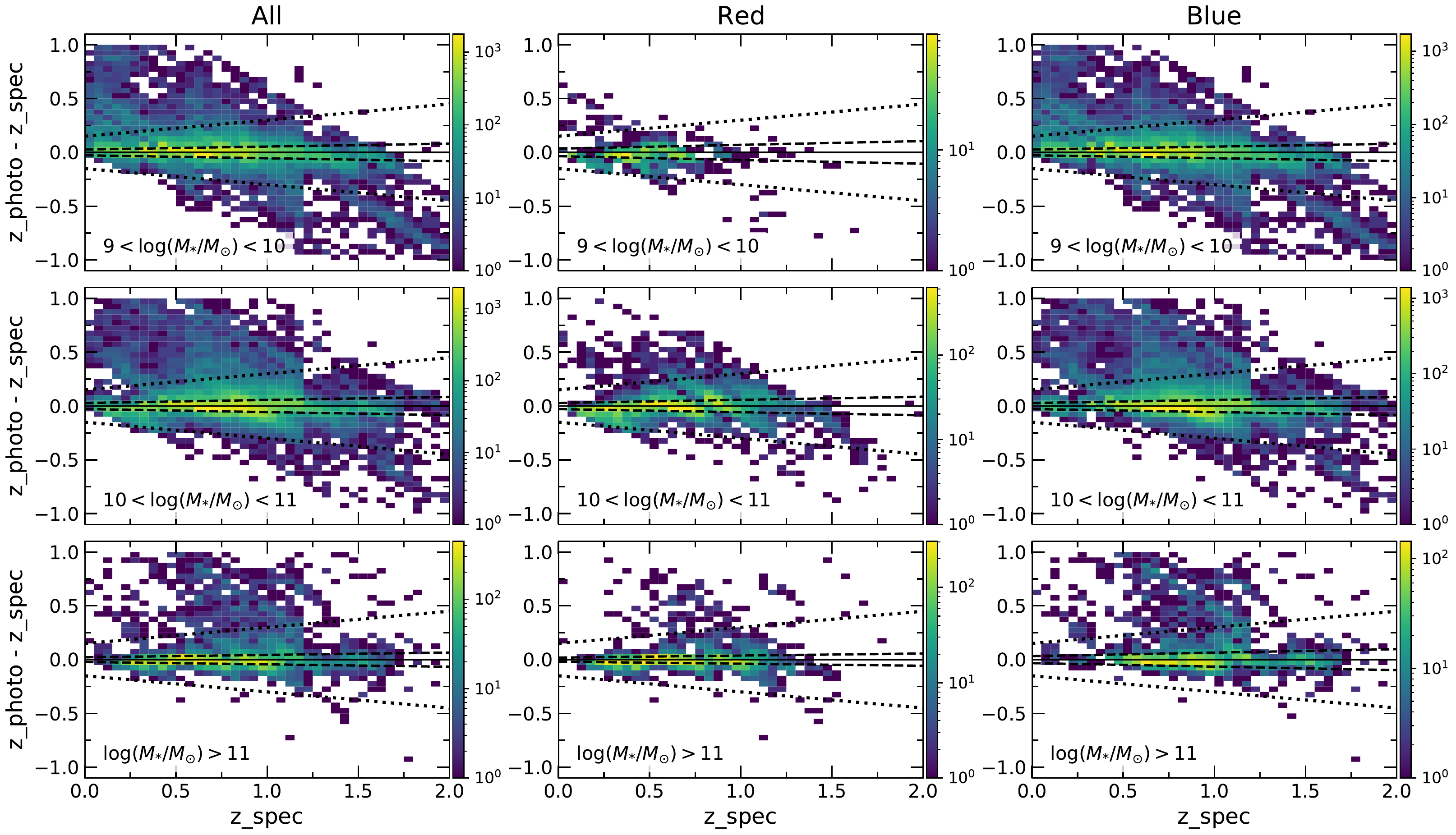}
    \caption{Comparison between $z_\mathrm{phot}$ and $z_\mathrm{spec}$. The left, middle and right columns correspond to all, red, and blue galaxies, respectively, and each row shows different stellar mass ranges. Solid and dashed lines show one-to-one relation and $\sigma_{z}(1+z_\mathrm{spec})$ ranges. Dotted lines are borders for outliers.}
    \label{fig:zphoto_zspec}
\end{figure*}

We show the statistical qualities of photo-\textit{z} estimates over the full redshift range (i.e., $0<z<6$) in Table~\ref{tab:photoz_prec_all}. In Fig.~\ref{fig:zphoto_zspec}, we show the distributions of differences between photo-\textit{z} and spec-\textit{z}. The left, middle and right columns correspond to all galaxies, RGs, and BGs, respectively, and each row shows a different stellar mass range.

\begin{table*}
\centering
    \caption{The photo-\textit{z} precision for the all redshift sample.}
    \label{tab:photoz_prec_all}
	\begin{tabular}{cccccccccccccccc}
		\hline
		& \multicolumn{4}{c}{all} & $\ $ & \multicolumn{4}{c}{red} & $\ $ & \multicolumn{4}{c}{blue}  \\
		\cline{2-5} \cline{7-10} \cline{12-15}
		mass & N & $\sigma_{z}$& $b_{z}$ & $\eta$ & & N & $\sigma_{z}$ & $b_{z}$ & $\eta$ & & N &$\sigma_{z}$& $b_{z}$ & $\eta$ \\
		$\log(M_{*}/M_{\odot})$ & \# & & & \% & & \# & & & \% & & \# & & & \% \\
		\hline
		$[9,10)$ & $43547$ & $0.027$ & $0.008$ & $6.7$ & & $1218$ & $0.035$ & $-0.021$ & $5.2$ & & $42329$ & $0.027$ & $0.009$ & $6.7$\\  [2pt]
		$[10,11)$ & $51534$ & $0.028$ & $-0.005$& $6.1$ & & $37471$ & $0.029$ & $-0.009$ & $2.8$ & & $14063$ &  $0.028$ & $-0.004$ & $7.3$\\  [2pt]
		$[11,\inf)$ & $10005$ & $0.022$ & $-0.009$ & $7.5$ & & $3719$ & $0.019$ & $-0.012$ & $2.2$ & & $6286$ & $0.033$ & $0.001$ & $16.4$\\  [2pt]
		\hline \hline
		& \multicolumn{4}{c}{all} & $\ $ & \multicolumn{4}{c}{QG} & $\ $ & \multicolumn{4}{c}{SFG}  \\
		\cline{2-5} \cline{7-10} \cline{12-15}
		mass & N & $\sigma_{z}$& $b_{z}$ & $\eta$ & & N & $\sigma_{z}$ & $b_{z}$ & $\eta$ & & N &$\sigma_{z}$& $b_{z}$ & $\eta$ \\
		$\log(M_{*}/M_{\odot})$ & \# & & & \% & & \# & & & \% & & \# & & & \% \\
		\hline
		$[9,10)$ & $12177$ & $0.025$ & $0.010$ & $4.5$ & & $426$ & $0.034$ & $-0.013$ & $6.1$ & & $11751$ & $0.024$ & $0.010$ & $4.4$ \\  [2pt]
		$[10,11)$ & $12762$ & $0.026$ & $-0.005$ & $3.8$ & & $3122$ & $0.026$ & $-0.001$ & $2.4$ & & $9640$ & $0.026$ & $-0.006$ & $4.3$\\  [2pt]
		$[11,\inf)$ & $1885$ & $0.022$ & $-0.007$ & $7.3$ & & $930$ & $0.018$ & $-0.012$ & $2.5$ & & $955$ & $0.028$ & $-0.0004$ & $11.9$\\  [2pt]
		\hline
	\end{tabular}
	\begin{tablenotes}[normal]
	\item \textit{Notes.} Same as Table~\ref{tab:photoz_prec}, but for all ($0<z<6$) galaxies provided by the HSC-SSP.
    \end{tablenotes}
\end{table*}

\section{Systematics in stellar mass estimate and robustness of results}
\label{app:robustness}
It has been reported that there is an offset and a scatter between the stellar masses estimated by \textsc{MIZUKI} with the HSC-SSP data and those of another multi-wavelength catalogue including near-infrared data \citep{Tanaka2015,Tanaka2018}. In addition, there is a more general problem that the stellar masses of SFGs derived by SED fitting are systematically underestimated \citep{Sorba2015,Sorba2018,Abdurrouf2018,Martinez-Garcia2017,Mosleh2020}. To check whether such uncertainties affect the results of this paper, we first compare the stellar masses of our sample to those of COSMOS2015 \citep{Laigle2016}.

\begin{figure*}
	\includegraphics[width=2\columnwidth]{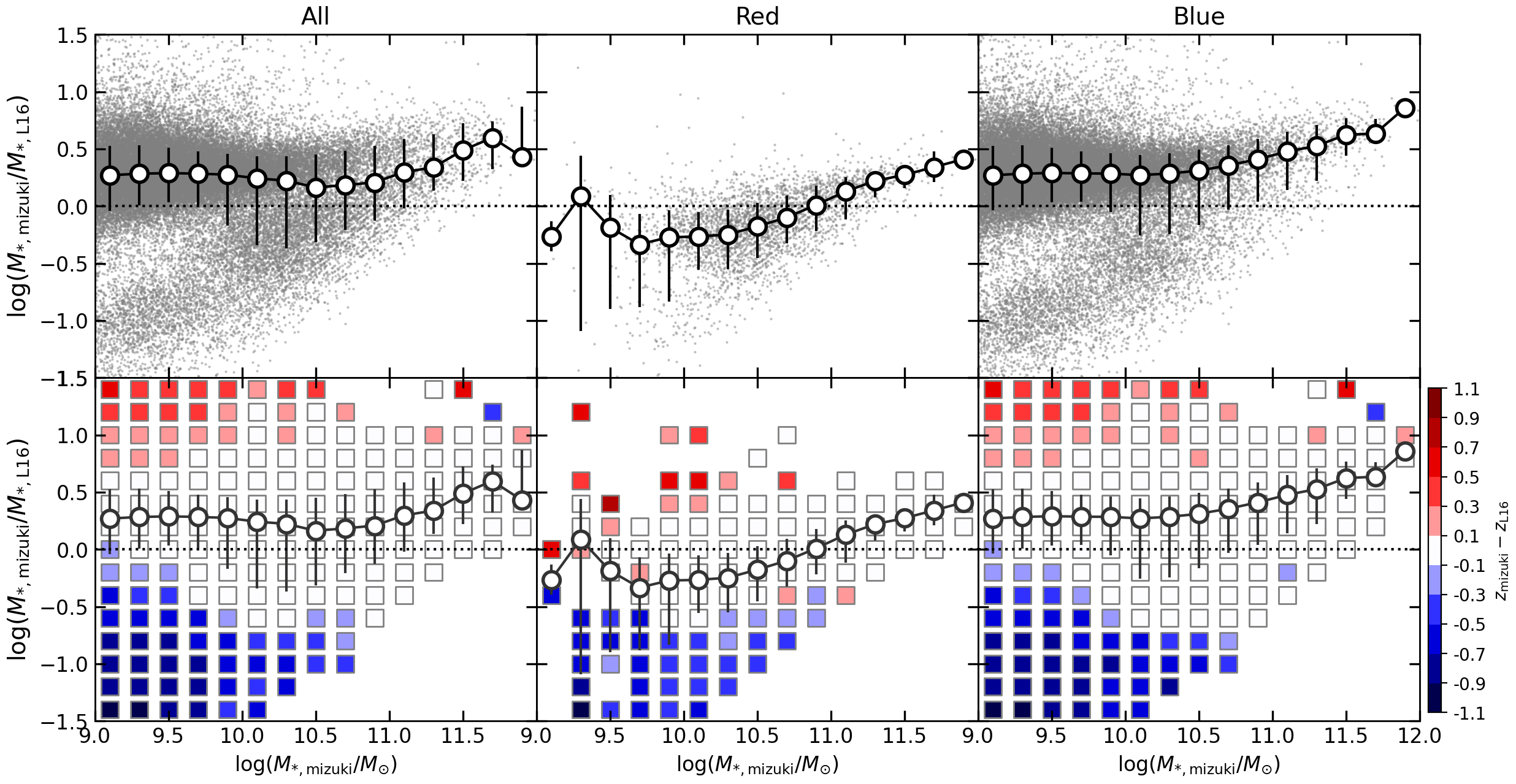}
    \caption{\textit{Top}: The ratio of estimated stellar masses between our sample and COSMOS2015 (\citealp{Laigle2016}, L16). The left, middle and right panels correspond to all, red, and blue galaxies, respectively. Grey dots are stellar mass ratios of individual galaxies. White points indicate median values of the ratio in bins at $0.2\,\mathrm{dex}$ intervals along the x-axis with 68th percentiles as error bars. \textit{Bottom}: The median redshift differences between our catalogue and COSMOS2015 on the same planes as the top panels. Red and blue colours indicate positive and negative offsets in redshift estimates, respectively. White points are the same as in the top panels.}
    \label{fig:mass_comp}
\end{figure*}

For the cross-matched subsample described in \S~\ref{sec:class}, we calculate the ratios between stellar masses estimated in the two catalogues as shown in the top panel of Fig.~\ref{fig:mass_comp}. Similar to \citet{Tanaka2015,Tanaka2018}, our sample has a positive offset of $0.2\text{--}0.5\,\mathrm{dex}$ compared to COSMOS2015. We also find that different galaxy classes have different values of offset and scatter. In particular, BGs show a sub-sequence at $\log(M_{*}/M_{\odot})\lesssim10.5$ with a large ($\lesssim-0.5$) offset. To check whether these stellar mass discrepancies between the two catalogues are due to uncertainties in redshift estimates, we plot the median differences between the two redshift catalogues on the bottom panels of Fig.~\ref{fig:mass_comp}. Red and blue colours indicate positive and negative offsets in redshift estimates, respectively. We find that galaxies with large stellar mass discrepancies, most of which are BGs in the sub-sequence, tend to have large ($\gtrsim0.3$) redshift differences, suggesting that the large stellar mass discrepancies are mainly due to wrong photo-\textit{z} estimates in our sample. On the other hand, most of our sample has redshifts consistent with COSMOS2015. Therefore, the positive stellar mass offset is likely due to other systematics inherent in the SED fitting codes.

Given these possible systematics in stellar mass estimates, we test the robustness of the observed RFE and conformity. First, regarding the stellar masses of COSMOS2015 as reference values, we calculate the correction factor for galaxy number counts at a given stellar mass bin as:
\begin{equation}
\label{eq:ms_corr}
    C(M_{*,i})=\frac{N_\mathrm{L16}(M_{*,i})}{N_\mathrm{mizuki}(M_{*,i})},
\end{equation}
where $N_\mathrm{L16}(M_{*,i})$ and $N_\mathrm{mizuki}(M_{*,i})$ are the numbers of galaxies in the given stellar mass bin, $M_{*,i}$, when we use L16's and \textsc{MIZUKI}'s stellar mass estimates, respectively. Then, we calculate the SMFs and the red fractions applying this correction factor to the number counts of field and satellite galaxies in the four DUD fields. Here, we use the same central galaxies as in the main body of this paper since massive galaxies in the \textsc{MIZUKI} catalogue are also relatively massive in COSMOS2015.

\begin{figure*}
	\includegraphics[width=2\columnwidth]{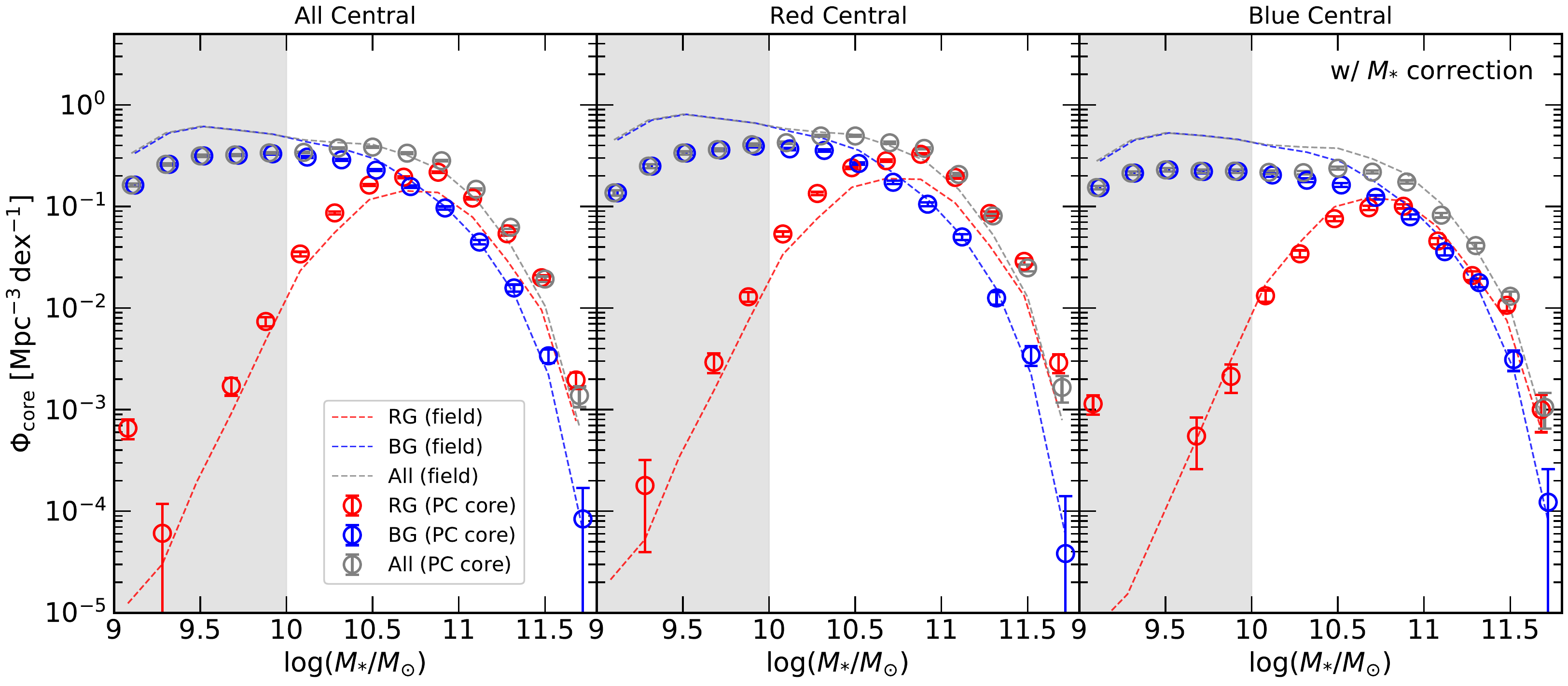}
    \caption{The SMFs of galaxies in PC cores with correction of stellar mass distribution defined by equation~\eqref{eq:ms_corr}. The meaning of the panels and the symbols are the same as Fig.~\ref{fig:SMF}.}
    \label{fig:smf_masscorr}
\end{figure*}

\begin{figure*}
	\includegraphics[width=2\columnwidth]{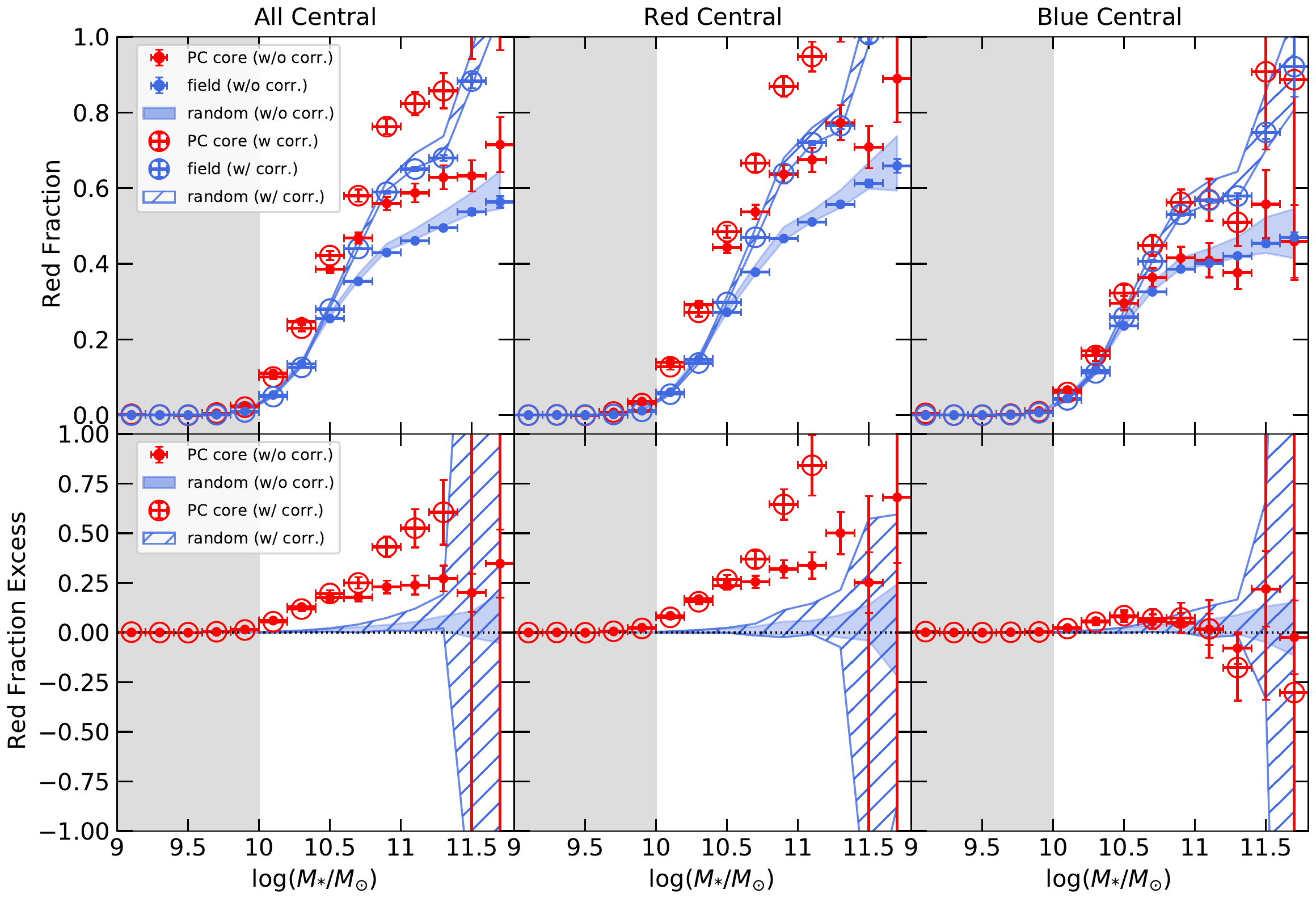}
    \caption{\textit{Top}: The red galaxy fraction $f_\mathrm{r}$ in PC cores (red open circles) and the field (blue open circles) with the correction of stellar mass distribution, overplotted with the original values shown in Fig.~\ref{fig:fr}. Blue hatches are the 68th percentiles of the $f_\mathrm{r}$ distribution measured around random points with the correction. Note that the range of the vertical axis is different from that of Fig.~\ref{fig:fr}.
    \textit{Bottom}: The RFEs in PC cores with the correction (open circles). The 68th percentiles of RFEs measured around random points are shown as hatches. Other symbols are the same as the bottom panel of Fig.~\ref{fig:fr}.
    }
    \label{fig:fr_masscorr}
\end{figure*}

\begin{figure}
	\includegraphics[width=\columnwidth]{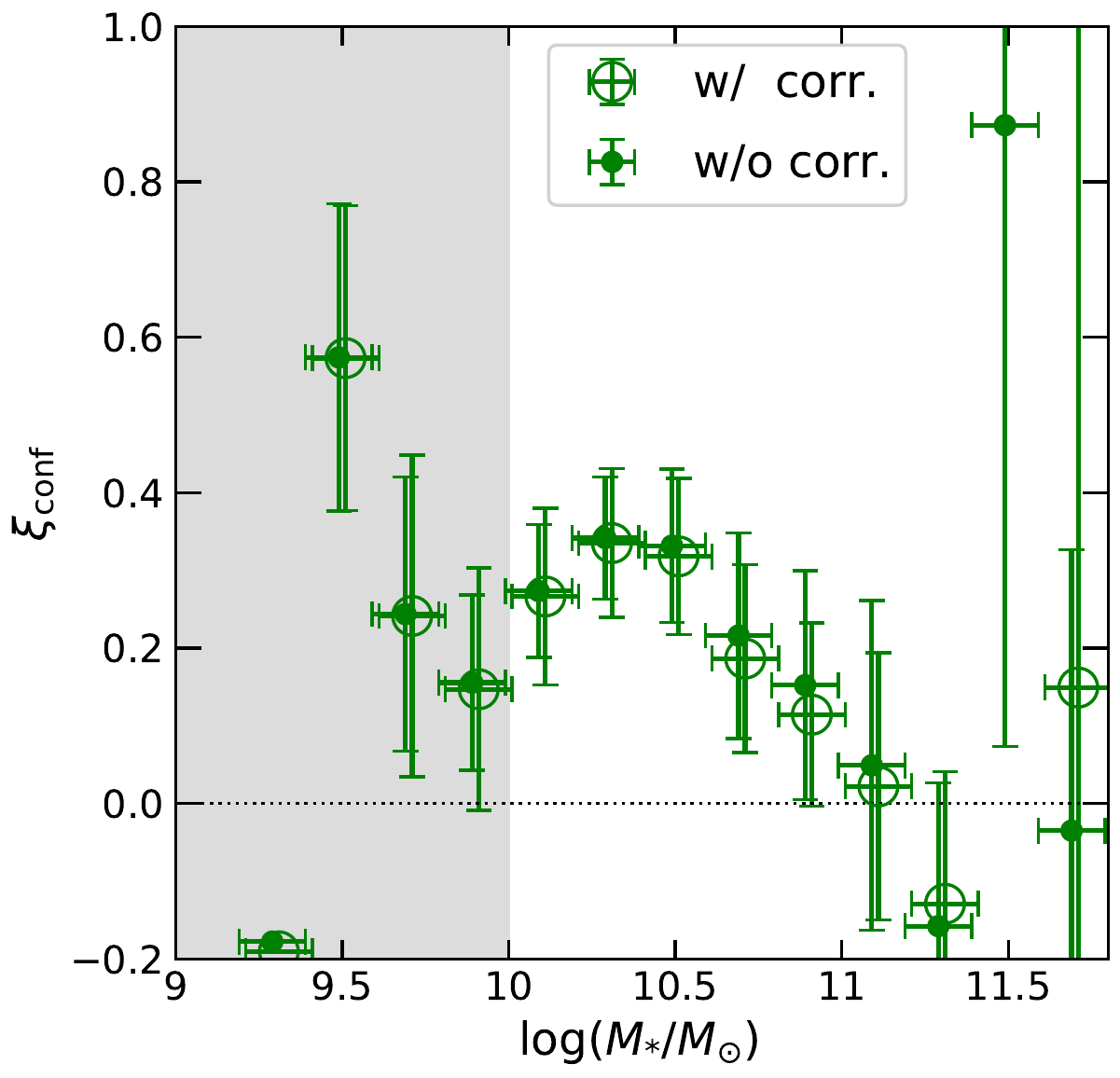}
    \caption{The strength of conformity, $\xi_\mathrm{conf}$, with (open circles) and without (dots) the correction of stellar mass distribution. Data points are slightly offset along the horizontal axis for clarity. These two are almost identical in all stellar mass bins.}
    \label{fig:xi_conf_masscorr}
\end{figure}

The corrected SMFs are shown in Fig.~\ref{fig:smf_masscorr}. Since the correction factor of equation~\eqref{eq:ms_corr} is cancelled when we calculate the ratios of SMFs between PC cores and the field, this correction does not change the difference in the shapes of the SMFs between PC cores and the field. Therefore, the discussion in \S~\ref{sec:smf} is unchanged.
With the correction, the absolute values of red fractions and RFEs become higher as shown in Fig.~\ref{fig:fr_masscorr}. However, their increasing trends with stellar mass are not changed. The RFE around blue centrals is within the uncertainties of the red fraction of field galaxies even after the correction. Hence, galactic conformity still exists. Indeed, as shown in Fig.~\ref{fig:xi_conf_masscorr}, $\xi_\mathrm{conf}$ before and after the correction are almost identical. 

Here, we consider another uncertainty in the stellar mass estimate. In general, spatially unresolved SED fitting may miss some fraction of stellar mass in SFGs due to the `outshining' effect: the light from bright young stars masks the light from faint and old low-mass stars that contain a significant fraction of the galaxy's stellar mass \citep{Sorba2015,Sorba2018}. \citet{Sorba2018} have found that this effect depends on the galaxy's specific-SFR (SFR divided by stellar mass) and proposed correction factors (see their equation~(6)). We find that the correction factor for our BGs is $\lesssim 0.1\,\mathrm{dex}$ and that applying this correction does not change the main results in this paper.

Based on these facts, we conclude that the main results of this paper are at least qualitatively robust against uncertainties in stellar mass estimates, and the observed conformity is not artificial.

\bsp	
\label{lastpage}
\end{document}